\documentclass{article}

\usepackage[T1]{fontenc}
\usepackage[utf8]{inputenc}
\usepackage[margin=1in,top=1in,bottom=3cm]{geometry}
\usepackage{perpage}
\MakePerPage{footnote}
\usepackage{epsfig}
\usepackage{latexsym}
\usepackage{amsmath}
\usepackage{mathrsfs}
\usepackage{yfonts}
\usepackage{makeidx}
\usepackage{graphicx}
\usepackage{slashed}
\usepackage{hyperref}
\usepackage[titletoc]{appendix}
\usepackage{chngcntr}
\usepackage{bm}
\usepackage{multicol}
\usepackage{nameref}
\usepackage{textcomp}
\usepackage{changepage}
\usepackage{lettrine}
\usepackage{enumitem}
\usepackage[sortcites,backend=bibtex,style=numeric,sorting=anyvt,doi=false,url=false,giveninits=true,isbn=false]{biblatex}
\AtEveryBibitem{\clearfield{month}}
\AtEveryBibitem{\clearfield{day}}
\usepackage{tikzsymbols}
\hypersetup{citecolor=blue,colorlinks=true,linkcolor=blue,pdfstartview={FitH},linktoc=page,pdftitle={Maxwell-Decay-on-DSRN-BH.pdf}pdfauthor={Mokdad Mokdad}}
\usepackage{amsthm,amssymb}
\usepackage{subcaption}
\newtheorem{thm1}{Theorem}
\newtheorem{prop1}[thm1]{Proposition}
\newtheorem{lem1}[thm1]{Lemma}

\newtheoremstyle{TheoremNum}
        {\topsep}{\topsep}              
        {\itshape}                      
        {}                              
        {\bfseries}                     
        {.}                             
        { }                             
        {\thmname{#1}\thmnote{ \bfseries #3}}
    \theoremstyle{TheoremNum}

     \newtheoremstyle{TheoremNum}
        {\topsep}{\topsep}              
        {\itshape}                      
        {}                              
        {\bfseries}                     
        {.}                             
        { }                             
        {\thmname{#1}\thmnote{ \bfseries #3}\thmnumber{}}
    \theoremstyle{TheoremNum}

\renewcommand{\d}{\mathrm{d}}
\newcommand{\scri}{{\mathscr I}}

\newcommand{\R}{\mathbb{R}}
\newcommand{\dl}{\partial}

\newcommand{\hf}{\frac{1}{2}}

\newcommand{\lie}{\mathcal{L}}
\newcommand{\hook}{{\setlength{\unitlength}{11pt}   
                   \begin{picture}(.833,.8)
                   \put(.15,.08){\line(1,0){.35}}
                   \put(.5,.08){\line(0,1){.5}}
                   \end{picture}}}
               
\makeindex

\title{Conformal Scattering of Maxwell fields on Reissner-Nordstrøm-de Sitter Black Hole Spacetimes}
\author{Mokdad Mokdad\\LMBA -- Université de Bretagne Occidentale}

\begin{document}

\maketitle

\begin{abstract}
	\emph{We construct a complete conformal scattering theory for Maxwell fields in the static exterior region of a Reissner-Nordstrøm-de Sitter black bole spacetime. This is done
	using uniform energy decay results that we obtain in a separate paper \cite{mokdad_decay_2017}, to show that the trace operators are injective and have closed ranges.
	We then solve the Goursat problem (characteristic Cauchy problem) for Maxwell fields on the null boundaries showing that the trace operators are also surjective.}
\end{abstract}

\tableofcontents

\section{Introduction}

In the classic experiment of scattering one has a field  propagating in a medium with an obstacle~; an incoming plane wave hits the obstacle and scatters away from it as a superposition of outgoing plane waves. {Scattering theory}\index{Scattering theory} is a way of summarizing this evolution, which may involve complicated intermediate interactions of the field, described as the solution to an evolution equation, by constructing the map that, to the asymptotic behaviour of the solution in the distant past (incoming wave), associates its asymptotic behaviour in the distant future (outgoing wave). This can be done provided the asymptotic behaviour characterizes the solution completely. Radar systems make use of this characterization of the solution by its asymptotic profile to gain information about the medium and the obstacles it contains. This reconstruction is the aim of {inverse scattering}\index{Inverse scattering}.

\subsection{Analytic Scattering: Brief literature Overview}

Scattering theory proved to be a useful tool in the framework of general relativity to study the asymptotic influence of the geometry of spacetime on fields. Although in this current work we do not use an analytic approach to scattering, we very briefly touch on the history of the subject because this is part of the origin of conformal scattering and it helps to understand what new features the conformal approach bring to the domain. Scattering theory in black holes spacetimes played an essential role in the rigorous description of phenomena like superradiance, the Hawking effect, and quasi-normal modes (resonances of black holes which are related to gravitational waves). In 1980 S. Chandrasekhar \cite{chandrasekhar_mathematical_1984} used the stationary approach, resorting to a Fourier transformation in time, to study  quasi-linear modes of black hole spacetimes such as Schwarzschild, Reissner-Nordstr{\o}m, and Kerr. Chandrasekhar’s work systematically used the Newman-Penrose formalism to develop stationary scattering theories described in terms of the scattering matrix of transmission and reflection coefficients. And around the same time, M. Reed and B. Simon published ``\emph{Scattering Theory}'' the third volume of their classic series \cite{reed_methods_1972}. Then time-dependent scattering (based on the comparison of dynamics) of classical and quantum fields on the exterior of a Schwarzschild black hole were first studied by J. Dimock in 1985 \cite{dimock_scattering_1985} and by J. Dimock and B. Kay in 1986 and 1987 \cite{dimock_scattering_1986, dimock_classical_1986, dimock_classical_1987}. And in the 1990's, A. Bachelot produced an important series of papers starting with scattering theories for classical fields, Maxwell in 1990 and 1991 \cite{bachelot_gravitational_1991, bachelot_operateur_1990}, Klein-Gordon in 1994 \cite{bachelot_asymptotic_1994} and on the Hawking effect for a spherical gravitational collapse in 1997 \cite{bachelot_scattering_1997}, 1999 \cite{bachelot_hawking_1999} and 2000 \cite{bachelot_creation_2000}. J.-P. Nicolas in 1995 developed a scattering theory for classical massless Dirac fields \cite{nicolas_scattering_1995}, and a work on a non linear Klein-Gordon equation on the Schwarzschild metric (and other similar geometries) with partial scattering results obtained by conformal methods in 1995 \cite{nicolas_non_1995}. W.M. Jin in 1998 contributed to the subject with a construction of wave operators in the massive case \cite{jin_scattering_1998}, and F. Melnyk in 2003 obtained a complete scattering for massive charged Dirac fields \cite{melnyk_scattering_2003} and the Hawking effect for charged, massive spin-1/2 fields \cite{melnyk_hawking_2003}. In 1999 I. Laba and A. Soffer \cite{laba_global_2000} obtained complete scattering for the nonlinear Schr\"{o}dinger equation on Schwarzschild manifolds. Then people started using commutator methods and Mourre theory. This led to scattering theories on the Kerr metric. One paper appeared in 1992 due to S. De Bièvre, P. Hislop and I.M. Sigal \cite{debievre_scattering_1992} on scattering theory for the wave equation on non-compact manifolds by means of a Mourre estimate. A complete scattering theory for the wave equation, on stationary, asymptotically flat space-times, was subsequently obtained by D. H\"{a}fner in 2001 using the Mourre theory \cite{hafner_completude_2001}. Time-dependent scattering theories on Kerr black holes were obtained by D. H\"{a}fner in 2003 \cite{hafner_sur_2003} and in 2004 by D. H\"{a}fner and J.-P. Nicolas for massless Dirac fields using a Mourre estimate \cite{hafner_scattering_2004}. In 2005 T. Daudé produced scattering theories for Dirac fields in various spacetimes \cite{daude_propagation_2004, daude_scattering_2004, daude_scattering_2005},  and in 2010 he published results on time-dependent scattering for charged Dirac \cite{daude_time-dependent_2010}, before moving to several works on inverse scattering in general relativity. In 2014 M. Dafermos, I. Rodnianski, and Y. Shlapentokh-Rothman developed scattering theory for the scalar wave equation on Kerr exterior backgrounds in the subextremal case \cite{dafermos_scattering_2014}. 

\subsection{Conformal Scattering}

In the present work, we construct a {Conformal Scattering}\index{Conformal Scattering} theory. Conformal scattering is a geometrical approach to time-dependent scattering based on Penrose {conformal compactification}\index{Conformal compactification}: a rescaling of the metric and the fields using conformal factors. This enables the definition of a {scattering operator}\index{Scattering operator}, the fundamental object in the theory. This operator associates to the asymptotic behaviour of the solution in the distant past, its asymptotic behaviour in the distant future. The asymptotics of the solution are the scattering data and are given as restrictions of the conformally rescaled solution on past and future null infinities and are called {radiation fields}\index{Radiation fields}. With suitable energy estimates, which is a crucial step in the theory, the scattering data completely characterizes the solution. This can be viewed as a characteristic Cauchy problem, also called a {Goursat problem}. This is an initial-value problem where data is given at null infinity instead of some spacelike hypersurface as in the non-characteristic case. The resolution of the Goursat problem is in the core of conformal scattering theory.

\subsubsection*{The Main Ingredients}

\begin{figure}
	\centering
	\includegraphics[scale=1]{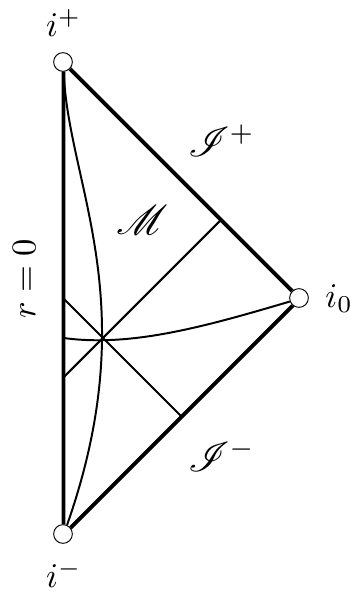}
	\caption{\emph{Penrose diagram of $\hat{\mathscr{M}}$ the conformal compactification of Minkowski spacetime $\mathscr{M}$ with timelike, spacelike, and null curves.}}
	\label{fig:ConfCompMinko}
\end{figure}
We describe the essential steps of the general strategy of conformal scattering.

\begin{description}
	\item[Conformal compactification.] In the words of R. Penrose, conformal compactification is a technique to ``make infinity finite''. A {globally hyperbolic spacetime}\index{Globally hyperbolic spacetime}\footnote{A spacetime that admits a spacelike hypersurface that intersect every inextendible causal curve exactly once.} $(\mathscr{M},g)$, with suitable asymptotic structure, such as asymptotic flatness, is rescaled and replaced by an ``unphysical'' Lorentzian manifold with boundary $(\hat{\mathscr{M}},\hat{g})$. $\hat{\mathscr{M}}$ is called the {conformal compactification}\index{Conformal compactification} of $\mathscr{M}$, with $\partial \hat{\mathscr{M}}=\scri$ representing points at infinity of $(\mathscr{M},g)$, and int$\hat{\mathscr{M}}=\mathscr{M}$. The new metric is {conformally related}\index{Conformally related metrics} to the original metric by $$\hat{g}=\Omega^2 g\; ,$$ for an appropriate choice of a smooth non-negative boundary function $\Omega$ defined on $\hat{\mathscr{M}}$. $\Omega$ called the {conformal factor}\index{Conformal factor}. It is positive on $\mathscr{M}$ and becomes zero on $\scri$, the asymptotic regions where $g$ becomes infinite, and $\d \Omega |_\scri \ne 0$ (figure \ref{fig:ConfCompMinko}). What is important is to define things in a way such that the new metric has some differentiability on the boundary hypersurface $\scri$. Now, the asymptotics of $\mathscr{M}$ can be studied using local techniques on $\hat{\mathscr{M}}$, without resorting to complicated limit arguments when studying, for example, the radiation fields of a physical field on the original spacetime. A {conformally invariant equation}\index{Conformally invariant equation} is an equation defined on $\mathscr{M}$ for $g$ such that whenever $\Phi$ is a solution to the equation, then for some $s\in\R$, the {rescaled field}\index{Rescaled field}\footnote{See \cite{penrose_spinors_1987} for the precise definition.} $\hat{\Phi}:=\Omega^{s}\Phi$ is a solution to the same equation but defined on $\hat{\mathscr{M}}$ for the rescaled metric $\hat{g}$. Examples of conformally invariant equations are the conformal wave equations, Dirac equation, and Maxwell's equations. Working with this class of equations that admit such rather explicit transformation law under conformal rescaling ensures that we can study the equation on the rescaled spacetime and gain information on its behaviour in the physical spacetime. Conformal scattering theories have been obtained on generic non-stationary spacetimes \cite{mason_conformal_2004, joudioux_probleme_2010}, but let us here assume the existence of a global Killing timelike (causal) vector field $\tau$ for simplicity. As the just cited works illustrate, this symmetry assumption can be relaxed to more general situations such as {asymptotically simple spacetimes}\index{Asymptotically simple spacetime} defined in \cite{chrusciel_existence_2002, chrusciel_mapping_2003,corvino_scalar_2000, corvino_asymptotics_2006}. We note that not all spacetimes admit a conformal compactification with the needed regularity of the rescaled metric at the boundary. This is in fact related to the decay of the Weyl curvature at infinity. When the required compactification exists, different parts of the boundary will correspond to different ways of going to infinity (along spacelike, timelike, or null curves). Also, in the cases of black holes, part or all of the conformal boundary will be the horizon or horizons. Horizons are finite null hypersurfaces for the physical metric and when the whole conformal boundary is made of horizons, conformal rescalings are not required~; even in such a case we talk about conformal scattering because we use the same approach based on the resolution of a Goursat problem at the null boundary. We note that such cases are more amenable to extending the method to non-conformally invariant equations since there is no conformal rescaling involved. For more details on the topic of conformal rescaling and compactification we refer to \cite{penrose_asymptotic_1963, penrose_conformal_1964, penrose_zero_1965, penrose_spinors_1987, penrose_spinors_1988}.

	\item[Cauchy problem: Defining the trace operators.] The scattering operator is defined using two operators called the past and the future {trace operators}\index{Trace operator} $T^\pm$. The past trace operator associates to data at some finite instant of time ($t=0$), data in the infinite past ($t=-\infty$). The future trace operator is defined similarly. These operators are defined between a normed energy space $\mathcal{H}$\index{$\mathcal{H}$} on a Cauchy hypersurface of the compactified spacetime $\hat{\mathscr{M}}$ and a normed energy spaces $\mathcal{H}^\pm$\index{$\mathcal{H}^\pm$} on the boundary parts $\scri^\pm$. The energy norms are defined by contracting the timelike or causal vector field $\tau$ with the stress-energy tensor $\mathbf{T}$ of the studied equations in order to define the energy current $J_a=\tau^b\mathbf{T}_{ab}$, and the norm is then the energy flux across the considered hypersurface: $$\mathcal{E}_{\tau,\Sigma_0}=\int_{\Sigma_0} \tau^a \mathbf{T}_{ab} \d \sigma^b \qquad\mathrm{and}\qquad \mathcal{E}_{\tau,\scri^\pm}=\int_{\scri^\pm} \tau^a \mathbf{T}_{ab} \d \hat{\sigma}^b\; .$$ 
	The general construction of the future operator goes as follows: For a given finite energy data $\hat{\Phi}_0$ on the spacelike Cauchy hypersurface $\Sigma_0$, we solve the Cauchy problem on $\hat{\mathscr{M}}$ to get a solution $\hat{\Phi}$ of our equations. The future radiation field, or the image of $\hat{\Phi}_0$ by the future trace operator, is then the trace (a restriction) of the solution $\hat{\Phi}$ to the future boundary $\scri^+$, i.e. $T^+(\hat{\Phi}_0)=\hat{\Phi}|_{\scri^+}$. The past trace operator is defined similarly (figure \ref{fig:traceoperatorsintro}). Of course, not all constructions follow this exact steps. Depending on the asymptotic structure of the spacetime and the equations we are studying, some intermediate steps may be required, and the definition of the trace operator may differ slightly. For example, while the above scheme generally describes the situation of the wave equation on Minkowski spacetime, additional steps are needed for different spacetimes depending on the nature of the timelike and and spacelike infinities, $i^\pm$ and $i_0$ (see \cite{nicolas_conformal_2016}). On the other hand, the trace operators for Maxwell's equations do not associate to the initial Cauchy data the full restriction of the field, but rather a part of it. This is because of the constraint equations that should be satisfied by the solution to the evolution problem. This is the case we treat in this work. For other situations we refer for example to \cite{mason_conformal_2004, nicolas_conformal_2016, joudioux_conformal_2012}.
\end{description} 

Let us for the sake of this general overview assume that the studied equations are linear, this entails that the trace operators themselves are linear operators, yet, this is not an absolute necessity for the construction of a conformal scattering theory, see \cite{joudioux_probleme_2010} for example.

\begin{description}[style =sameline] 
	
	\begin{figure}
		\centering
		\begin{minipage}{.5\textwidth}
			\centering
			\includegraphics[scale=1]{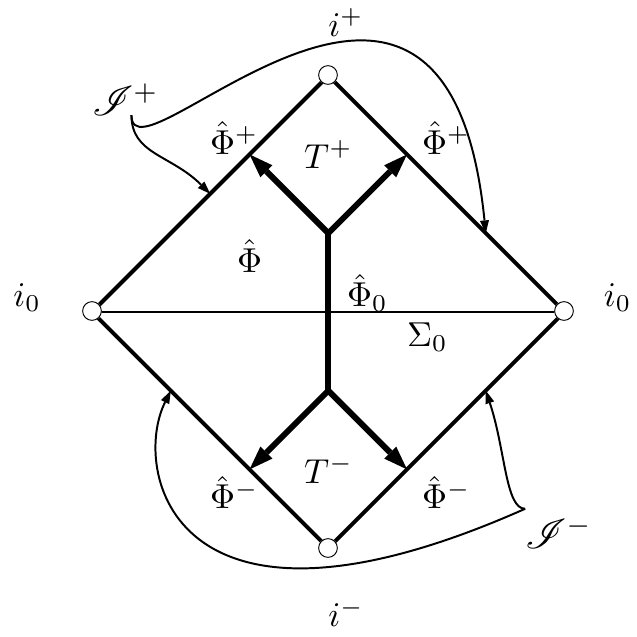}
			\caption{\emph{The trace operators $T^\pm$.}}
			\label{fig:traceoperatorsintro}
		\end{minipage}%
		\begin{minipage}{.5\textwidth}
			\centering
			\includegraphics[scale=1]{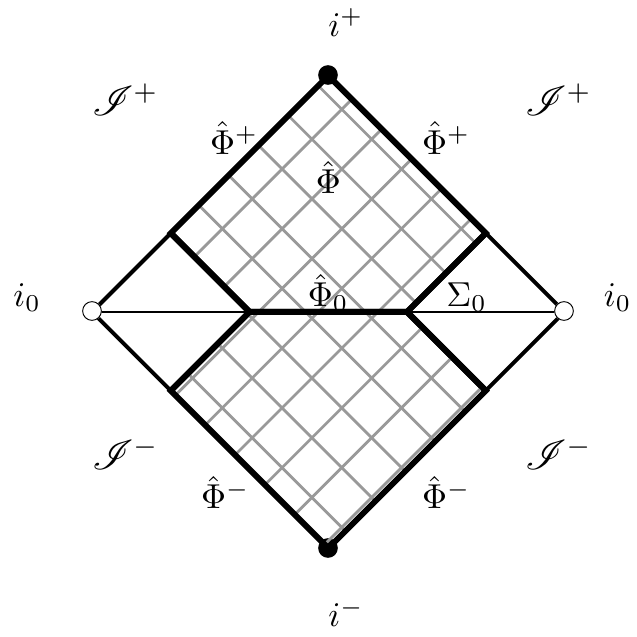}
			\caption{\emph{The case of regular $i^\pm$.}}
			\label{fig:energyestimatesintrononsingulari+}
		\end{minipage}
	\end{figure}
	
	\item[Energy estimates: The trace operators are one-to-one and have closed ranges.] The next step is to show that the trace operators are bijective. In fact, the above construction of the trace operators is usually done first for a dense subset of the finite energy space $\mathcal{H}$ on $\Sigma_0$ such as smooth compactly supported functions. If one proves uniform energy estimates both ways between the initial Cauchy data in the dense subset and their images under the trace operator, then the operator extends to the whole of $\mathcal{H}$ as a one-to-one map with a closed range. In some cases, one can prove exact energy identities, and the trace operators preserve the energy norms in this case, i.e. they are partial isometries. Ways of getting the uniform estimates depend on the structure of the spacetime at infinity and the properties of the stress-energy tensor. If the stress-energy tensor of the original unrescaled equations is divergence-free i.e. conserved, and conformally invariant, as for the Maxwell's equations, then working with the rescaled quantities $\hat{\Phi}$ and $\hat{g}$ has the important advantage of seeing all the involved hypersurfaces as regular hypersurfaces at finite distances, in particular $\scri^\pm$. If we are on Minkowski spacetime, a simple application of Stokes' theorem, or more precisely the divergence theorem, yields the required energy identities: $$\mathcal{E}_{\tau,\Sigma_0}= \mathcal{E}_{\tau,\scri^\pm}.$$ Even if the rescaled metric is singular at $i_0$, as long as the initial data is supported away from $i_0$, finite propagation speed guarantees that the solution does not see the singular spacelike infinity since it is zero in a neighbourhood of it, and the above technique can be applied without essential modification thanks to the density of compactly supported functions in the energy space (figure \ref{fig:energyestimatesintrononsingulari+}). In the case of black hole spacetimes, timelike infinities are singular. This constitutes an important difficulty and finite propagation speed will not help us here since the singularity lies in the future of any initial data no matter how small its compact support may be. What we need is a suitable decay of the solutions near timelike infinities so that we can rule out the accumulation of energy at these singularities. In such situations the estimates can be obtained as follows. We consider an achronal hypersurface $S_s$ ($s>0$) for the rescaled metric that forms a regular closed hypersurface with the future boundary $\scri^+$ and $\Sigma_0$ \footnote{Except possibly for $i_0$ when it is singular, but the compact support keeps us from running into troubles there.} as shown in figure \ref{fig:closedsurfacesintro}. Becuase $\tau$ is Killing, the divergence theorem now implies that
	$$\mathcal{E}_{\tau,\Sigma_0} = \mathcal{E}_{\tau,\scri^+_s}+\mathcal{E}_{\tau,S_s}.$$
	\begin{figure}
		\centering
		\includegraphics[scale=1]{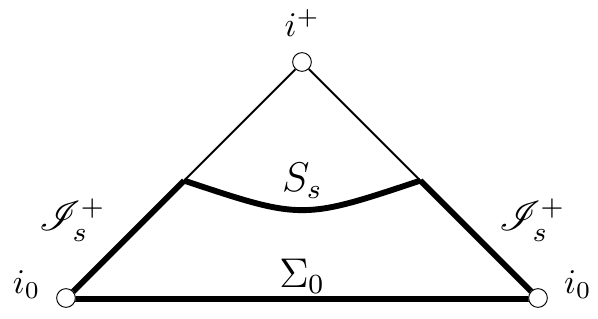}
		\caption{\emph{The closed hypersurfaces of the compactified spacetime.}}
		\label{fig:closedsurfacesintro}
	\end{figure}
	Assume that $S_s$ accumulates on $i^+$ as $s \rightarrow +\infty$. Here is where the decay is needed, namely to show that
	$$\lim_{s\rightarrow +\infty}\mathcal{E}_{\tau,S_s}=0\; ,$$
	and the conservation law follows: $$\mathcal{E}_{\tau,\Sigma_0}= \mathcal{E}_{\tau,\scri^+}.$$ Clearly, the same can be done in the past direction. Obtaining the desired decay is usually a separate problem that has its difficulties. This is partly why we proved the decay results of \cite{mokdad_decay_2017}. In a different setting, such as the wave equation on the Schwarzschild metric, the energy estimates are not as direct since the stress-energy tensor is not conformally invariant, and hence the stress-energy tensor of the rescaled equation is not conserved. However, it happens that one can recover the conservation law for the wave equation on Schwarzschild black hole spacetimes since the error term is a divergence \cite{nicolas_conformal_2016}, here too a decay result \cite{dafermos_lectures_2008} is needed to ensure no information is lost at the singular $i^\pm$. The current decay results use techniques that require local information that are too precise for a scattering theory. It is however not clear yet what are the minimal decay assumptions needed for conformal scattering.

	\item[Goursat problem: The trace operators are onto.]\index{Goursat problem} The third and last step in defining the scattering operator is to prove that the trace operators we defined are surjective, this comes down to solving the Goursat problem on a null hypersurface for data in dense subsets of the finite energy spaces $\mathcal{H}^\pm$, usually smooth and compactly supported functions. This means that we need to find for a given smooth compactly supported Goursat data, say $\hat{\Phi}^+$ on $\scri^+$, a $\hat{\Phi}_0\in \mathcal{H}$ such that $T^+(\hat{\Phi}_0)=\hat{\Phi}^+$. Taking into account the well-posedness of the Cauchy problem, we need then to find a finite energy solution to the equations that has $\hat{\Phi}^+$ as its trace on $\scri^+$. One way of solving the characteristic Cauchy problem is to approach the null conformal boundary by spacelike hypersurfaces. Goursat data are projected as part of the Cauchy data on the spacelike slices by means of congruences of null geodesics in the neighbourhood of $\scri$. The solution to the Goursat problem is then obtained using uniform energy estimates, weak convergence, and compactness methods \cite{mason_conformal_2004, hormander_remark_1990}. In some case, some ``reversible'' modifications to the setting is needed before applying the methods just mentioned or the results they produce. For example, one can still apply the results of \cite{hormander_remark_1990} where spatial compactness is needed, to spacetimes that are not spatially compact by a cut-extend construction that transports the problem into a framework suitable for \cite{hormander_remark_1990}. This is done in section \ref{goursatproblemsection} following the construction done in \cite{nicolas_conformal_2016}, but there the situation is needs a step more to the singularity at $i_0$.
	
	\item[Scattering operator]\index{Scattering operator} With the Goursat problem solved, the trace operators $T^\pm$ become isometries between the boundary energy spaces $\mathcal{H}^\pm$ on $\scri^\pm$ and the initial energy space $\mathcal{H}$ on $\Sigma_0$. We can then define the scattering operator $S:\mathcal{H}^-\rightarrow\mathcal{H}^+$ by $S=T^+\circ (T^-)^{-1}$ and it is an isometry. Although this construction of the scattering operator relies on a choice of Cauchy hypersurface used to construct the trace operators $T^\pm$, however, the scattering operator maps the past radiation fields to the future radiation fields independently of the choice of the intermediate spacelike hypersurface and the theory is in fact truly covariant as Penrose hinted in \cite{penrose_zero_1965}.
\end{description} 

\subsubsection*{History}
The introduction of ``points at infinity'' in a consistent way where these points constitute a hypersurface boundary $\scri$ to a manifold whose interior is conformally identical with the original space-time, was first done by R. Penrose around 1964 \cite{penrose_zero_1965,penrose_conformal_1964, penrose_asymptotic_1963} and presented in his classic book with W. Rindler \cite{penrose_spinors_1988, penrose_spinors_1987} in the 1980's. This idea was first motivated by the fact that massless free-field equations are conformally invariant if interpreted in a suitable way, so their behaviour at ``infinity'' can be studied at this hypersurface. In the same period of early 1960's F.G. Friedlander introduced his notion of {radiation fields}\index{Radiation field}  \cite{friedlander_radiation_1962, friedlander_radiation_1964, friedlander_radiation_1967}: In spherical coordinates, a radiation field of a solution $u(t,r,\omega)$ to the wave equation is a function $v(t,\omega)$ on $\R \times \mathcal{S}^2$ given by the limit 
$$v(t,\omega)=\lim_{r\rightarrow +\infty} ru(t+r,r,\omega)\; .$$ 
Penrose in \cite{penrose_zero_1965} explicitly states that scattering is a motivation for introducing the conformal compactification technique: ``The technique affords a covariant approach to the definition of radiation fields in general relativity.'' Meanwhile, P.D. Lax and S.R. Phillips developed their theory of scattering  \cite{lax_scattering_1967} in 1967. The Lax-Phillips scattering theory for the wave equation on flat spacetime is based on a translation representative of the solution which is reinterpreted as an asymptotic profile of the field along outgoing null geodesics, analogous to Friedlander's radiation field. Fifteen years after Penrose discussed radiations fields in the conformal setting, Friedlander saw the connection between Lax-Phillips theory of scattering and his notion of radiation fields, and in 1980  the first actual conformal scattering theory appeared in his founding paper \cite{friedlander_radiation_1980}. The paper treated the case of the conformal wave equation in a static asymptotically flat spacetime with a fast enough decay at infinity to ensure a smooth conformal compactification including at spacelike and timelike infinities. The principle of the construction was first to reinterpret the scattering theory as the well-posedness of the Goursat problem for the rescaled equation at null infinity, then to solve this Goursat problem. Friedlander as well as J.C. Baez, I.E. Segal and Zhou Z.F. who pushed his ideas further in 1989-1990 \cite{baez_global_1989, baez_global_1990, baez_conserved_1990, baez_scattering_1989, baez_scattering_1989-1} worked exclusively on static backgrounds. Right after \cite{baez_global_1990}, L. H\"{o}rmander solved the Goursat problem for a wave equation on generic null hypersurfaces in a spatially compact spacetime \cite{hormander_remark_1990}. With this, and knowing that constructing conformal scattering theories amounts to solving a Goursat problem on a compactified spacetime, the road to non-stationary spacetimes was clear. Still, no one pushed it in this direction until 2004 when L. Mason and J.P. Nicolas picked up Friedlander's ideas and applied them to scalar waves\footnote{The result on waves was completed in another paper in 2009 \cite{mason_regularity_2009} by the same authors.}, Dirac, and Maxwell fields on generically non-stationary asymptotically simple spacetimes \cite{mason_conformal_2004}. J. Joudioux in 2012 \cite{joudioux_conformal_2012} constructed a conformal scattering theory for a non-linear wave equation on non-stationary backgrounds. And in 2013 J.P. Nicolas produced a paper \cite{nicolas_conformal_2016} on a conformal scattering theory for the wave equation on Schwarzschild black holes. In these recent works, \cite{mason_conformal_2004, nicolas_conformal_2016, joudioux_conformal_2012} and the current work, the resolution of the Goursat problem is based on methods following the work of H\"{o}rmander \cite{hormander_remark_1990} which deal with the Goursat problem using energy estimates for the wave equation, weak convergence, and compactness. The data in \cite{hormander_remark_1990} is given on a general weakly spacelike Lipschitz hypersurface (including null), then the problem is solved by changing the equation using a parameter in front of the Laplacian in the wave equation to slow down the propagation speed so that the given weakly spacelike Lipschitz hypersurface becomes spacelike for to the modified equation\footnote{The resolution of the Goursat problem on a Lipschitz spacelike hypersurface is done by approximation with smooth spacelike hypersurfaces then using the well-posedness of the Cauchy problem on them.}. While in \cite{mason_conformal_2004} the energy estimates of \cite{hormander_remark_1990} are used, the authors, instead of slowing down the propagation speed, approach null infinity by spacelike hypersurfaces without changing the equation. Here in our work we directly apply \cite{hormander_remark_1990} to show that the Goursat problem for Maxwell fields on Reissner-Nordstr{\o}m-de Sitter black holes is well-posed.

The ultimate purpose of conformal scattering is to use conformal methods to construct scattering theories, not to reinterpret existing scattering theories in conformal terms. The idea of replacing spectral analysis by conformal geometry is the door to the extension of scattering theories to general non-stationary situations, which may be inaccessible to spectral methods. In \cite{nicolas_conformal_2016, mason_conformal_2004, friedlander_radiation_1980}, the reinterpretation is done in addition to the conformal construction, giving more insight on questions such as the required decay for a conformal scattering theory, or whether a conformal scattering theory and a scattering theory defined in terms of wave operators are equivalent or not: Some spectral scattering theories cannot be reinterpreted as conformal scattering, but when the spacetime has the right asymptotic structure and the equation considered is conformally invariant, the question is valid. For the time being, the methods used require these two conditions, however it is interesting to know whether and how they can be extended to more general situations of conformally non-invariant equations which include the massive cases. The setting in the present paper may be suitable to construct conformal scattering for massive fields\footnote{See the end of section \ref{sec:workdone}.}.

\subsection*{Work Done}\label{sec:workdone}

Here, we address the topic of conformal scattering on the exterior region of RNdS black holes, and construct a scattering operator establishing the isometric correspondence between null data on past horizons and null data on future horizons of the static exterior region of RNDS in the case of three horizons. The paper is divided as follows.  

\paragraph{Section \ref{sec:geometricframework}:} We discuss the set--up of the work. Notations and tools required are introduced. Also,  some of the properties of the RNDS spacetime in the case of three horizons and its maximal extension are reviewed.

\paragraph{Section \ref{sec:traceoper}:} In this section we construct the trace operators and show that they are injective and norm preserving after establishing conservation laws. We start the section by expressing the Maxwell field in null tetrad formalisms adapted to the geometry of our spacetime. 
We next define the energy spaces on the horizons associated to the smoothly extended vector field $T$ given by $\dl_t$ on the static exterior region, and thus specify the Goursat data.
By the decay results on achronal hypersurfaces that we obtained in \cite{mokdad_decay_2017}, we establish an
energy identity or a conservation law between data on the initial Cauchy hypersurface $\Sigma_0 $ $(t=0)$ and data on the horizons. Thereby, the energy of the Cauchy data is equal to the energy of the Goursat data. 
The global hyperbolicity of the spacetime guarantees the well-posedness of the Cauchy problem and allows us to define each trace operator  between  the
space of finite energy constrained Cauchy data as a partial isometry (an isometry into its range) into the space of finite energy Goursat data. 
Showing that the trace operators are invertible, i.e. isometries between the full spaces of finite energy, requires solving the Goursat problem on the horizons which we do in this section. 
For this, we proceed as follows. Since the spin components of the Maxwell field satisfies a system of coupled wave equations.
This allows us to transform the problem from Maxwell's equations to wave equations. 
We can with a simple construction adapt the setting to the framework of H\"{o}rmander's results in \cite{hormander_remark_1990} that prove the well-posedness of the Goursat problem for a general class wave equation. 
This gives the existence of the solution to our system of wave equations.
The next step is to reinterpret the solution of this wave equations as a Maxwell field. The main idea of the proof is to use the fact that one can go back and forth from Maxwell's equations (with perturbations) to wave equations by successive applications of the Maxwell operator. This allow us to obtain a system of wave equations whose well-posedness (again by \cite{hormander_remark_1990}) entails the required interpretation of the solution to the wave Goursat problem as a Maxwell field.

It is worth mentioning that the conformal scattering we construct here is done without conformal compactification! This is because scattering data is taken on the horizons which are regular null hypersurfaces for the original metric on the maximal extension of RNdS black hole. Nevertheless, the results we obtain can be applied to any spherically symmetric spacetime satisfying the conditions stated in \cite{mokdad_decay_2017} with a conformal compactification when needed, the rest goes through essentially without modifications since Maxwell's equations are conformally invariant and in fact the rescaled Maxwell field tensor is equal to the unrescaled one, and the stress-energy tensor is also conformally invariant.

\section{Geometric Framework}\label{sec:geometricframework}

We start by recalling the Reissner-Nordstr{\o}m-de Sitter metric. 

\subsection{Reissner-Nordstr{\o}m-de Sitter Spacetime}

One of the spherically symmetric solutions of Einstein-Maxwell Field equations in the presence of a positive cosmological constant is the Reissner-Nordstr{\o}m-de Sitter solution (RNDS). It models a non-rotating spherically symmetric charged black hole with mass and a charge, in a de Sitter background. The de Sitter background means that there is a cosmological horizon beyond which lies a dynamic region that stretches to infinity. While the Reissner-Nordstr{\o}m nature entails that near the singularity, depending on the relation between the mass and the charge, one has a succession of static and dynamic regions separated by horizons. The properties of this spacetime that we summarize here, are detailed in \cite{mokdad_reissner-nordstrom-sitter_2017}. 

The Reissner-Nordstr{\o}m-de Sitter metric is given in spherical coordinates by 
\begin{equation}\label{RNdSmetric}
g_\mathcal{M}=f(r)\d t^2-\frac{1}{f(r)}\texttt{d}r^2-r^2\d \omega^2,
\end{equation}
where
\begin{equation}\label{f(r)}
f(r)=1-\frac{2M}{r}+\frac{Q^2}{r^2}-\Lambda r^2 \; ,
\end{equation}
and $\d \omega^2$ is the Euclidean metric on the $2$-Sphere, $\mathcal{S}^2$, which in spherical coordinates is,
\begin{equation*}
\d \omega^2=\d \theta^2 + \sin(\theta)^2\d \varphi^2 \; ,
\end{equation*}
and $g_\mathcal{M}$ is defined on \index{$\mathcal{M}$}$\mathcal{M}=\R_t \times ]0,+\infty[_r \times \mathcal{S}^2_{\theta,\varphi}$ . Here $M$ is the mass of the black hole, $Q$ is its charge, and $\Lambda$ is the cosmological constant. We assume that $Q$ is real and non zero, and $M$ and $\Lambda$ are positive.

The metric in these coordinates appear to have singularities at $r=0$ and at the zeros of $f$. Only the singularity at $r=0$ is a real geometric singularity at which the curvature blows up. The apparent singularities at the zeros of $f$ are artificial and due to this particular choice of coordinates. The regions of spacetime where $f$ vanishes are essential features of the geometry of the black hole, they are the event horizons or {horizons}\index{Horizon} for short, and $f$ is called the {horizon function}\index{Horizon function}. If $f$ has three positive zeros and one negative, then the zeros in the positive range corresponds in an increasing order respectively to the {Cauchy horizon} or {inner horizon}\index{Inner horizon}\index{Cauchy horizon}, the {horizon of the black hole} \index{Horizon of the black hole} or the {outer horizon}\index{Outer horizon}, and the {cosmological horizon}\index{Cosmological horizon}. In this case, $f$ changes sign at each horizon and one has static and dynamic regions separated by these horizons.

In this work, we are interested in the construction of a conformal scattering theory for Maxwell fields on the RNDS spacetime in the case of three horizons. Precisely, on the closure of the static region between the horizon of the black hole and the cosmological horizon, which we refer to as the {exterior static region}\index{Exterior static region}. This part of the spacetime contains a photon sphere in This is a hypersurface where photons orbit outside the black hole. It consists of purely rotational null geodesics. The photon sphere is an important feature of the geometry that affects the decay of the Maxwell solutions by trapping them and a priori may cause a loss in the information for the scattering operator.
 In fact, under the the following conditions,
\begin{equation}\label{GC}
Q\neq0 \quad \textrm{and} \quad 0<\Lambda < \frac{1}{12Q^2} \quad \textrm{and} \quad M_1 < M < M_2 \; ,
\end{equation}
where 
\begin{gather}\label{naming}
R=\frac{1}{\sqrt{6\Lambda}} \quad;\quad \Delta=1-12Q^2\Lambda \quad ; \quad m_1=R\sqrt{1-\sqrt{\Delta}} \quad ;\quad m_2=R\sqrt{1+\sqrt{\Delta}}\\ M_1=m_1-2\Lambda m_1^3 \quad ; \quad M_2=m_2-2\Lambda m_2^3 \label{naming2}\; , 
\end{gather}
we have
\begin{prop1}[Three Positive Zeros and One Photon Sphere]\label{1photonsphere}
	The function $f$ has exactly three positive distinct zeros if and only if (\ref{GC}) holds. In this case, there is exactly one photon sphere in the static exterior region of the black hole defined by the portion between the largest two zeros of $f$.
\end{prop1}
\begin{proof}
	This is proved in \cite{mokdad_reissner-nordstrom-sitter_2017}.
\end{proof}

Consider the following open subsets of $\mathcal{M}$, which we also refer to by I, II, III, and IV, respectively:
\begin{eqnarray*}
	\mathrm{U}_1&=&\R_t \times ]0,r_1[_r \times \mathcal{S}^2_{\theta,\varphi}\; ; \\
	\mathrm{U}_2&=&\R_t \times ]r_1,r_2[_r \times \mathcal{S}^2_{\theta,\varphi}\; ; \\
	\mathrm{U}_3&=&\R_t \times ]r_2,r_3[_r \times \mathcal{S}^2_{\theta,\varphi}\; ; \\
	\mathrm{U}_4&=&\R_t \times ]r_3,+\infty[_r \times \mathcal{S}^2_{\theta,\varphi}\; ,
\end{eqnarray*}
and let $I_i$ be the corresponding interval of $r$ such that
\begin{equation}\label{intervalsofrI_i}
\mathrm{U}_i=\R_t \times I_i \times \mathcal{S}^2_{\theta,\varphi}\; .
\end{equation}

With the assumption of (\ref{GC}), let the zeros of $f$ be $r_0<0<r_1<r_2<r_3$. For $r>0$, we define the Regge-Wheeler coordinate function $r_*$ by requiring 
$$\frac{\d r_*}{\d r}=\frac{1}{f(r)}>0.$$
The Regge-Wheeler radial coordinate have the following expression:
\begin{equation*}
r_*(r)=\sum_{i=0}^3 a_i \ln |r-r_i| + a \quad ; \; a_i=-\frac{r_i^2}{\Lambda}\prod_{j\ne i}\frac{1}{(r_i-r_j)} \quad ; \; a=-\sum_{i=0}^3 a_i \ln |P_2-r_i|
\end{equation*}
where $$\left\{ r=P_2=\frac{3M+\sqrt{9M^2-8Q^2}}{2} \right\}$$ is the photon sphere hypersurface.

We now introduce the chart $(t,r_*,\theta,\varphi)$ over the exterior static region  $\mathcal{N}=\R_t \times]r_2,r_3[\times\mathcal{S}^2_{\omega}$. We see that $r_*$ is a strictly increasing continuous function of $r$ (thus a bijection) over the interval $]r_2,r_3[$, and ranges from $-\infty$ to $+\infty$. We also have $\partial_{r_*}=f\partial_r$ and $\d r = f \d r_*$. The RNdS metric in these coordinates is:
\begin{equation}\label{Reggewheelermetric}
g_\mathcal{N}=f(r)(\d t^2-\d r_*^2)-r^2\d \omega^2 \- .
\end{equation}

\begin{figure}
	\centering
	\includegraphics[scale=1]{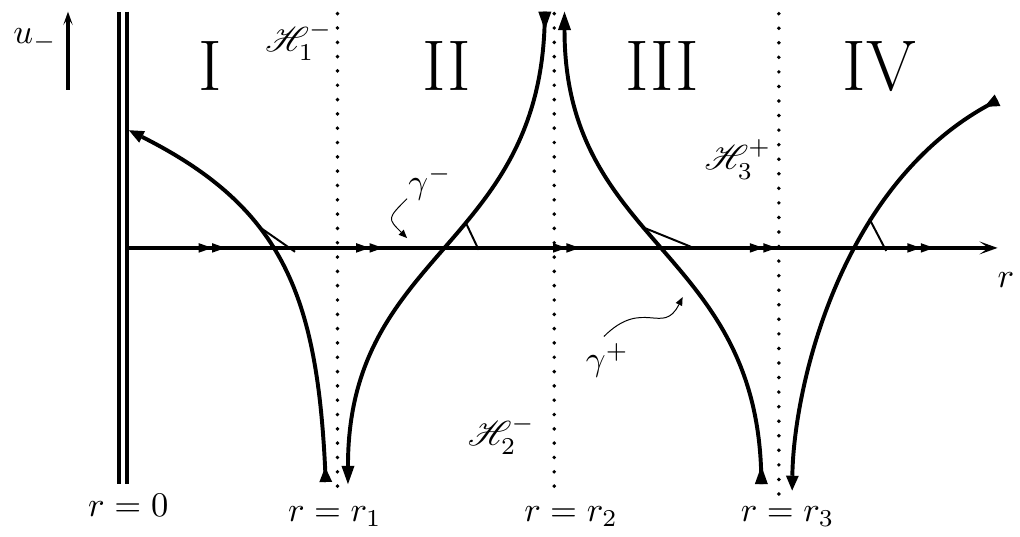}
	\caption{\emph{$\mathcal{M}^-_F$ and the integral curves of $Y^\mp$.}}
	\label{fig:REFcoord}
\end{figure}

To cover the boundaries of $\mathcal{N}$ we need to introduce other charts. The Eddington-Finkelstein retarded coordinate chart on $\mathrm{U}_i$ is 
$$({u_-}_i,r,\omega)\in \R_{{u_-}} \times {I_i}_r \times \mathcal{S}^2_{\omega}\;,$$ with ${u_-}_i=t-{r_*}_i$. In this chart the metric is: 
\begin{equation}\label{RetardedEddington-Finkelsteinmetric}
g=f(r)\d {u_-}_i^2 +2\d {u_-}_i \d r -r^2\d\omega^2\; ,
\end{equation} 
This expression of the metric is analytic for all values $(u_-,r,\omega)\in \R \times ]0,+\infty[ \times \mathcal{S}^2$, including $r=r_i$. The Lorentzian manifold $\mathcal{M}^-=\R_{u_-}\times ]0,+\infty[_r\times \mathcal{S}^2_{\omega}$\index{$\mathcal{M}^-$} with the metric (\ref{RetardedEddington-Finkelsteinmetric}) is called the \emph{retarded Eddington-Finkelstein extension}\index{Retarded extension} of the RNdS manifold. Taking the orientation of $\mathcal{M}$, $(\dl_{u_-},\dl_r,\dl_\theta,\dl_\varphi)$ is positively oriented on $\mathcal{M}^-$, and when $\dl_r$ is chosen to be future-oriented\footnote{This is \emph{not} the coordinate vector field $\dl_r$ of the chart $(t,r,\omega)$. If we denote the Eddington-Finkelstein retarded coordinates by $(u_-,r_-(=r),\omega)$ then $\dl_{r_-}=f^{-1}\dl_t + \dl_r=Y^-$.}, we denote $\mathcal{M}^-$ by $\mathcal{M}^-_F$\index{$\mathcal{M}^-_F$} and call it the {future retarded extension}\index{Future retarded extension} (figure \ref{fig:REFcoord}).  

\begin{figure}
	\centering
	\includegraphics[scale=0.85]{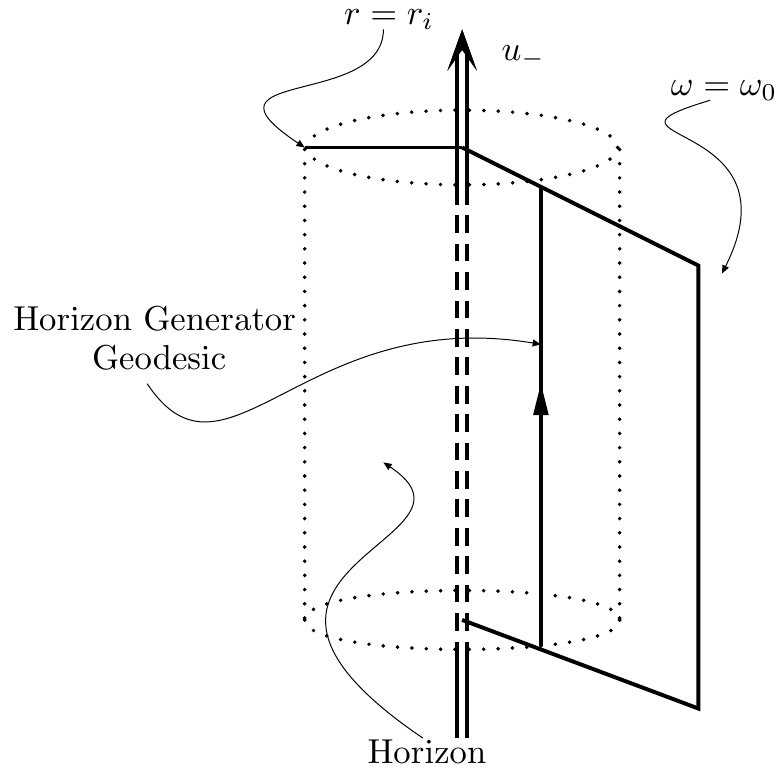}
	\caption{\emph{Integral curves of $e^{\hf f'(r_i)u_-}\dl_{u_-}$ at $r=r_i$ are null geodesics that generate the horizon $\{r=r_i\}$.}}
	\label{hrzngenerator}
\end{figure}

For an observer in III, light coming from the singularity and passing through the first two event horizons of the black hole is travelling forward in time and hence is from the past. Therefore the observer will consider the singularity to be in the past as well as the {past inner horizon}\index{Past inner horizon}\index{$\pm\mathscr{H}_i^\pm$} $\mathscr{H}_1^-=\R_{u_-}\times\{r=r_1\}\times \mathcal{S}^2_\omega$, and the {past outer horizon}\index{Past outer horizon} $\mathscr{H}_2^-=\R_{u_-}\times\{r=r_2\}\times \mathcal{S}^2_\omega$, which are now regular null hypersurfaces. Similarly, the observer can only send but never receive any signal from the last horizon and $\scri$. In this extension, we denote $\scri$ by \index{$\scri^+$}$\scri^+$ since it lies in the future of the observer, and so does the {future cosmological horizon}\index{Future cosmological horizon} $\mathscr{H}_3^+=\R_{u_-}\times\{r=r_3\}\times \mathcal{S}^2_\omega$ which is a regular null hypersurface for the metric (\ref{RetardedEddington-Finkelsteinmetric}). The null horizons are generated by null geodesics each lying in a fixed angular plane (figure \ref{hrzngenerator}). This means that at the horizon some ``photons hover'' in place at $r=r_i$ and $\omega=\omega_0$. 

We refer to $\mathcal{M}^-$ with the opposite time-orientation as $\mathcal{M}^-_P$ the {past retarded extension}\index{Past retarded extension}.
We can also define the {advanced Eddington-Finkelstein null coordinate}\index{Advanced coordinate} $u_+=t+r_*$ and a new extension $\mathcal{M}^+$ \index{$\mathcal{M}^+$} covered by a single chart $(u_+,r,\omega)\in \R_{u_+} \times ]0,+\infty[_r \times \mathcal{S}^2_\omega=\mathcal{M}^+$. It is endowed with the analytic metric
\begin{equation}\label{AdvancedEddington-Finkelsteinmetric}
g=f(r)\d {u_+}^2 -2\d {u_+} \d r -r^2\d\omega^2\; ,
\end{equation}
where $(\dl_{u_+},\dl_r,\dl_\theta,\dl_\varphi)$ is positively oriented and $-\dl_r$ is future-oriented. This is the {future advanced Eddington-Finkelstein extension}\index{Future advanced extension} $\mathcal{M}^+_F$\index{$\mathcal{M}^+_F$}. Similarly, with $\dl_r$  future-oriented we get the {past advanced Eddington-Finkelstein extension}\index{Past advanced extension} $\mathcal{M}^+_P$\index{$\mathcal{M}^+_P$}.
In $\mathcal{M}^+_F$, we have the {future inner horizon}\index{Future inner horizon}\index{$\pm\mathscr{H}_i^\pm$} $\mathscr{H}_1^+=\R_{u_+}\times\{r=r_1\}\times \mathcal{S}^2_\omega$, the {future outer horizon}\index{Future outer horizon} $\mathscr{H}_2^+=\R_{u_+}\times\{r=r_2\}\times \mathcal{S}^2_\omega$, and the {past cosmological horizon}\index{Past cosmological horizon} $\mathscr{H}_3^-=\R_{u_+}\times\{r=r_3\}\times \mathcal{S}^2_\omega$. For the past extensions, $\mathcal{M}^\pm_P$, $\scri$ will be $\scri^-$, and we denote the horizon by a minus sign when we want to specify: $-\mathscr{H}_i^\pm$.


To cover the bifurcation spheres $\mathcal{S}_i$ where the horizons from different charts but of the same $r=r_i$ value meet, we need to introduce the Kruskal-Szekeres Extensions. With three families $\{A_{k,l}, B_{k,l}, C_{k,l}\}$ of these extension we can cover $\mathcal{M}^*$ the maximal analytic extension of the RNDS manifold as shown in figure \ref{fig:OpensConnected1} below.
\begin{figure}
	\hspace{-0.5cm}
	\includegraphics[width=\textwidth]{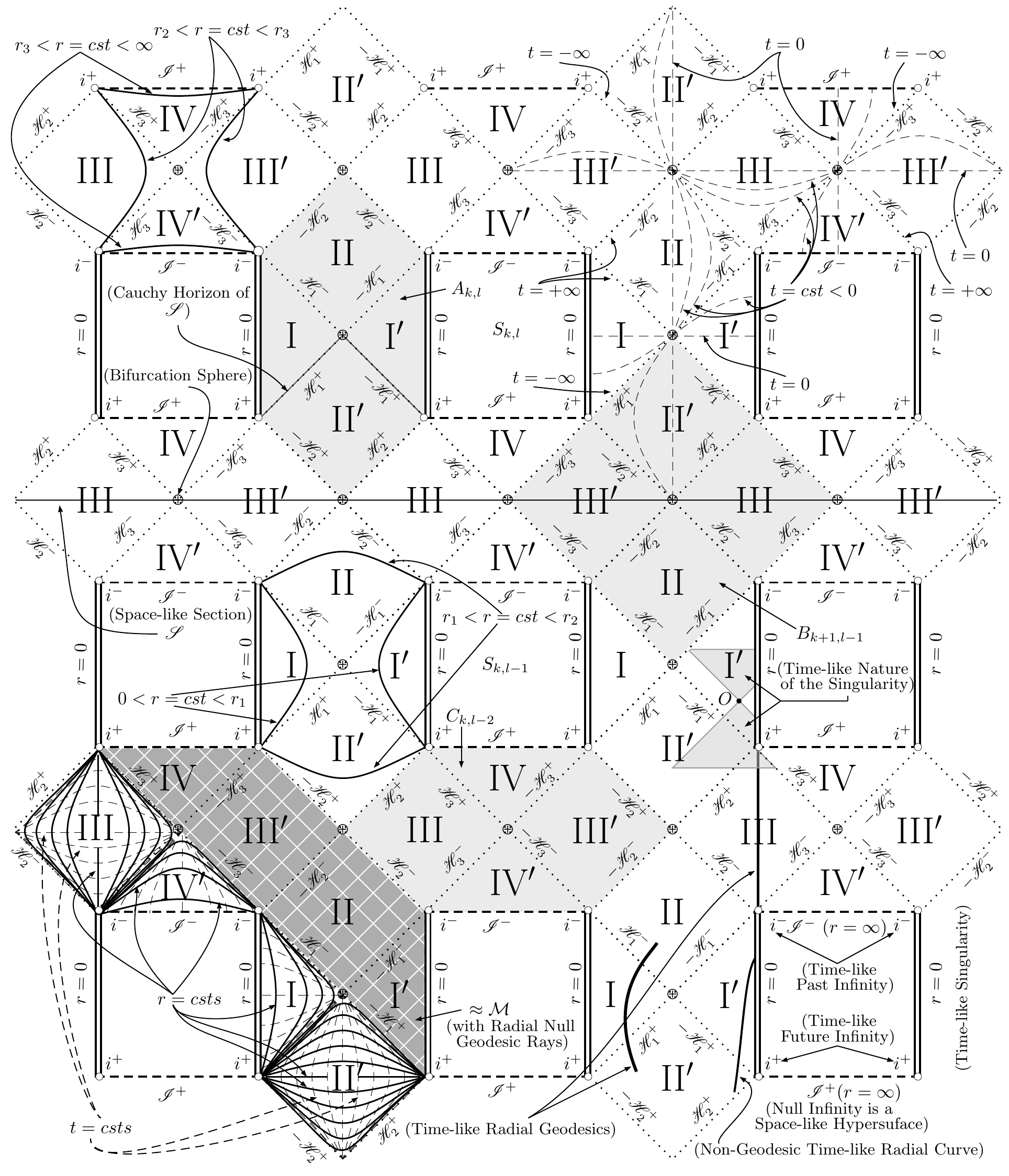}
	\caption{\emph{The Structure of $\mathcal{M}^\ast$.}}
	\label{fig:OpensConnected1}
\end{figure}

\subsection{Maxwell's Equations}

Let $F$ be a 2-form on the RNdS manifold $\mathcal{N}$. The source free {Maxwell's equations}\index{Maxwell's equations} can be written as
\begin{eqnarray}
\delta F&=&0 \; , \label{Maxeq1}\\
\d F&=& 0 \; ,\label{Maxeq2}
\end{eqnarray}
where $\delta=\star \d \star$, and $\star$ is the Hodge star operator. In abstract index notation,
\begin{eqnarray}
\nabla^a F_{ab} &=& 0 \; , \label{Maxeqabst1}\\
\nabla_{[ a} F_{bc ] }&=& 0 \; ,\label{Maxeqabst2}
\end{eqnarray}
and in coordinate form these translate to the following two sets of equations,
\begin{eqnarray}
g^{ac}\left(\partial_c F_{ab}-F_{db}\Gamma^d_{ca}-F_{ad}\Gamma^d_{cb}\right) &=& 0 \qquad \qquad\forall b,\label{Maxeqcoor1} \\
\partial_c F_{ab}+\partial_b F_{ca}+\partial_a F_{bc}&=& 0 \qquad \qquad \forall a,b,c. \label{Maxeqcoor2}
\end{eqnarray}
If taken in the coordinates $(t,r_*,\theta,\varphi)=(\tilde{x}^0,\tilde{x}^1,\tilde{x}^2,\tilde{x}^3)$, (\ref{Maxeqcoor1}) becomes respectively for $b=0,..,3$:
\begin{eqnarray}
&&\dl_1F_{10}+V\dl_2F_{20}+V\sin(\theta)^{-2}\dl_3F_{30}+2 rVF_{10}+VF_{20}\cot(\theta)+f'F_{01}=0, \label{Maxeqcoor10}\\
&&\dl_0F_{10}+V\dl_2F_{21}+V\sin(\theta)^{-2}\dl_3F_{31}+VF_{21}\cot(\theta)=0, \label{Maxeqcoor11} \\
&&\dl_1F_{12} + \dl_0F_{20}+V\sin(\theta)^{-2}\dl_3F_{32}=0, \label{Maxeqcoor12}\\
&&\dl_1F_{13} + \dl_0F_{30}+V\dl_2F_{23}+VF_{32}\cot(\theta)=0.\label{Maxeqcoor13}
\end{eqnarray}
where $V=fr^{-2}$\index{$V$}.

As much as equations (\ref{Maxeq1})-(\ref{Maxeqabst2}) are elegant and simple they are not the most convenient form for us to use in all arguments and calculations, and evidently neither are their expressions in coordinates. 
We instead use the tetrad formalism.
We use the components of the field in a general basis of the tangent space which might not be the canonical basis given by the coordinates. At each point, one defines a set of four vectors, called the tetrad, that forms a basis for the tangent space at that point. One  can then reformulate the field equations using this tetrad. In general relativity, it is natural to project on a null tetrad, which consists of two real null vectors and two conjugate null complex vectors usually defined as $X\pm iY$ for $X$ and $Y$ two spacelike real vectors. Here, we use a null tetrad on $\mathcal{N}$ given by two null real vectors and a two conjugate null complex vector tangent to the 2-Sphere $\mathcal{S}^2$:
\begin{eqnarray}\label{nulltetrad}
L &=& \dl_t +\dl_{r_*} \nonumber\\
N &=& \dl_t -\dl_{r_*} \nonumber\\
M &=& \dl_\theta +\frac{i}{\sin(\theta)}\dl_{\varphi} \\
\bar{M} &=&  \dl_\theta -\frac{i}{\sin(\theta)}\dl_{\varphi}\nonumber
\end{eqnarray}

We shall call this tetrad the ``{stationary tetrad}''. Using this tetrad, we can represent the Maxwell field by three complex scalar functions $\bm{\Phi}=(\Phi_{-1},\Phi_0,\Phi_1)$ called the {spin components}\index{Spin components} of the Maxwell field associated to the given tetrad, and defined by :
\begin{eqnarray}\label{spincomponentsPhis}
\Phi_1&=&F\left(L,M\right) \nonumber\\
\Phi_0 &=& \frac{1}{2}\left(V^{-1}F(L,N)+F\left(\bar{M},M\right)\right) \\
\Phi_{-1}&=&F\left(N,\bar{M}\right) \nonumber
\end{eqnarray}

We note that the tetrad we use, unlike those in the {Newman-Penrose formalism}\index{Newman-Penrose formalism}, are not normalized: A {normalized tetrad}\index{Normalized tetrad} is such that the inner product of the two null real vectors of the tetrad with each other equals $1$, and the product of the null complex  vector with its conjugate is $-1$, while all other products are zero. The formalism we use is a form of Geroch–Held–Penrose formalism (GHP), which does not require normalization\footnote{The conventional definition of spin components of an anti-symmetric tensor is slightly different. One normally defines it without the extra factor of $V^{-1}$ in the middle component $\Phi_0$. Here we carry on with the notation we used in \cite{mokdad_decay_2017} for obtaining decay. Also the usual way to label the components is different, conventionally, they are indexed by $0$, $1$, and $2$. The conformal weight and the spin weight are respectively related to the way the component change when we rescale the complex vector of the tetrad by a complex constant and the conjugate vector by the conjugate constant, and when rescaling the first null vector of the tetrad by a real constant and the second by the inverse constant. More precisely , the components transform as powers of the real rescaling constant, the power being the index of the component. For more on spin--components notations see \cite{penrose_spinors_1987,penrose_spinors_1988}.\label{footnotespinweight}}. The form of Maxwell's equations in this formalism is usually referred to as Maxwell's {compacted equations}\index{Compacted equations} (see \cite{penrose_spinors_1987}).

A straight forward coordinate calculation shows that in this framework, Maxwell's equations translate as follows.

\begin{lem1}[Maxwell Compacted Equations]\label{NewmanPenrose} $F$ satisfies Maxwell's equations (\ref{Maxeq1}) and (\ref{Maxeq2}) if and only if its spin components $(\Phi_1,\Phi_0,\Phi_{-1})$ in the stationary tetrad satisfy the compacted equations 
	\begin{eqnarray}
	N\Phi_1 &=& V M\Phi_0, \label{NewmanPenrose1} \\
	L\Phi_0 &=& \bar{M}_1\Phi_1, \label{NewmanPenrose2}\\
	N\Phi_0 &=& -M_1\Phi_{-1},  \label{NewmanPenrose3}\\
	L\Phi_{-1} &=& -V\bar{M}\Phi_0. \label{NewmanPenrose4}
	\end{eqnarray}
	where $M_1=M+\cot(\theta)$ and $\bar{M}_1$ is its conjugate.
\end{lem1}

\begin{figure}
	\centering
	\includegraphics[scale=0.8]{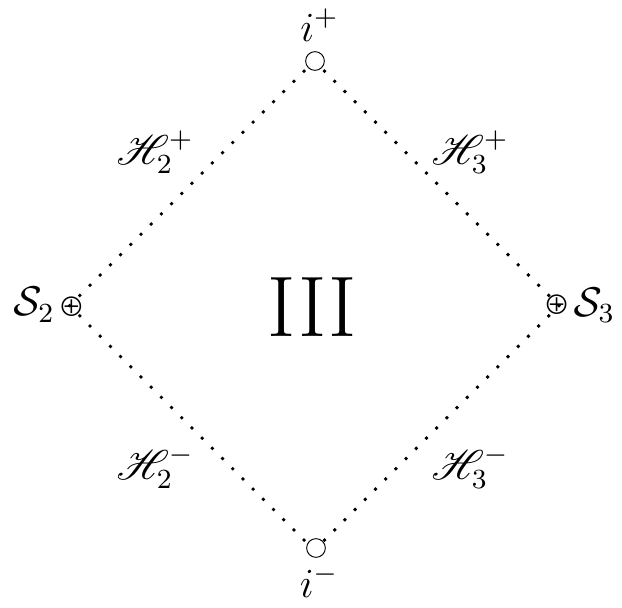}
	\caption{\textit{$\bar{\mathcal{N}}$: the closure of $\mathcal{N}$ in $\mathcal{M}^*$.}}
	\label{fig:closureofN}
\end{figure}

We need to study Maxwell fields up to the horizons, i.e. on 
\begin{equation}\label{barmathcalN}
\bar{\mathcal{N}}\; 
\end{equation}
\index{$\bar{\mathcal{N}}$} the closure of $\mathcal{N}$ in $\mathcal{M}^*$. The boundary of $\mathcal{N}$ consists of the future and past outer horizons $\mathscr{H}^\pm_2$, the future and past cosmological horizons $\mathscr{H}^\pm_3$, and the two bifurcation spheres $\mathcal{S}_2$ and $\mathcal{S}_3$. In addition we have the two singular timelike infinities $i^\pm$ which are not part of $\bar{\mathcal{N}}$ (figure \ref{fig:closureofN}). 

To do write Maxwell's compacted equations on the closure, we express Maxwell's equations in different null tetrads adapted to the geometry of the spacetime. Let $F$ be a Maxwell field on $\bar{\mathcal{N}}$. To write down its spin components we need to define a tetrad at each point of $\bar{\mathcal{N}}$. Simply extending the old tetrad $\{L,N,M,\bar{M}\}$ to the boundary will not work, particularly because one of the null vectors $L$ and $N$, which are given by $\dl_t\pm f\dl_r$ in $(t,r,\omega)$-coordinates, will always vanish on two of the horizons. For example, in $(u_-,r,\omega)$-coordinates, $L=f\dl_r$ and thus it vanishes on $\mathscr{H}^+_3$ and $\mathscr{H}^-_2$, which means that the tetrad $\{L,N,M,\bar{M}\}$ is singular there and does not form a basis of the tangent space. The same thing happens to $N$ on $\mathscr{H}^+_2$ and $\mathscr{H}^-_3$ as can be seen in the $(u_+,r,\omega)$-coordinates. However, if we rescale $L$ by the factor $f^{-1}$, the tetrad $\{\hat{L}=f^{-1}L=\dl_r,N=2\dl_{u_-}- f\dl_r,M,\bar{M}\}$ becomes a regular basis on $\mathcal{M}^-_F$, and in particular, on $\mathscr{H}^+_3$ and $\mathscr{H}^-_2$. We define the spin components of $F$ in this tetrad as:
\begin{eqnarray}\label{spincomponentsofFinLhat}
\hat{\Phi}_1&=&F\left(\hat{L},M\right) \nonumber\\
\Phi_0 &=& \frac{1}{2}\left(\hat{V}^{-1}F(\hat{L},N)+F\left(\bar{M},M\right)\right) \\
\Phi_{-1}&=&F\left(N,\bar{M}\right) \nonumber
\end{eqnarray} 
where $\hat{V}=f^{-1}V=r^{-2}$. Since the integral curves of $\hat{L}$ are the outgoing radial null geodesics, we refer to this tetrad as the {outgoing tetrad}\index{Outgoing tetrad}, and the spin components as the {outgoing components}\index{Outgoing spin components}. Compared to (\ref{spincomponentsPhis}), the components in the stationary tetrad $\{L,N,M,\bar{M}\}$, we have $\hat{\Phi}_1=f^{-1}{\Phi}_1$ while the other two components stay the same, hence we denote them by the same letters. The {incoming tetrad}\index{Incoming tetrad}\index{Incoming spin components} $\{L,\hat{N}=f^{-1}N,M,\bar{M}\}$ and the associated components are defined similarly. Neither of these two new tetrads defines a frame on the entire spacetime $\bar{\mathcal{N}}$, however, we can use the two tetrads with a partition of unity subordinate to the open sets $\mathcal{M}^-_F$ and $\mathcal{M}^+_F$ to define a tetrad that extends to all horizons. Using the relation $\hat{\Phi}_1=f^{-1}{\Phi}_1$ with the other components being the same, it is readily found that Maxwell compacted equations (\ref{NewmanPenrose}) in the outgoing tetrad take the following form:
\begin{eqnarray}
N\hat{\Phi}_1 &=& \hat{V} M\Phi_0 + f' \hat{\Phi}_1, \label{CEsoutgoing1} \\
\hat{L}\Phi_0 &=& \bar{M}_1\hat{\Phi}_1, \label{CEsoutgoing2}\\
N\Phi_0 &=& -M_1\Phi_{-1},  \label{CEsoutgoing3}\\
\hat{L}\Phi_{-1} &=& -\hat{V}\bar{M}\Phi_0. \label{CEsoutgoing4}
\end{eqnarray}
Similar equations for the incoming tetrad hold.

We know that if the Maxwell field is given as an exterior derivative of some $1$-form, one can then define a Lagrangian. By varying the $1$-form, the Euler-Lagrange equations will be Maxwell's equations. Using this Lagrangian, it is possible to define an energy-momentum tensor which by the Euler-Lagrange equations is divergence-free. In this case, if $F$ is a Maxwell field, then the energy-momentum tensor will be
\begin{equation}\label{energymomentumtensormaxwell}
\mathbf{T}_{ab}=\frac{1}{4} g_{ab}F^{cd}F_{cd}-F_{ac}{F_b}^c  \; .
\end{equation}
However, in general not all Maxwell fields admit a global potential. But,  one can use the same expression of energy-momentum tensor as in the case where the field is an exact form. By direct calculations, we see that it is still divergence-free if Maxwell's equations are satisfied.

We define the energy flux of a Maxwell field across a hypersurface $\Sigma_{t}=\{t=cst\}$ by
\begin{equation*}
E_T[F](t)=\frac{1}{4}\int_{\Sigma_t} |\Phi_1|^2 +2V|\Phi_0|^2 + |\Phi_{-1}|^2 ~\d r_* \d^2 \omega \; .  
\end{equation*}
This norm is the natural energy associated with Maxwell's equations and it can be defined geometrically: Consider an energy-momentum tensor $\mathbf{T}_{ab}$ that is a $(0,2)$-symmetric tensor i.e. $\mathbf{T}_{ab}=\mathbf{T}_{(ab)}$, and which is divergence-free i.e. $\nabla^a\mathbf{T}_{ab}=0$. Let $X$ be a vector field and  ${}^{(X)}\pi_{ab}=\nabla_a X_b + \nabla_b X_a$ be its deformation tensor. If $\mathcal{U}$ is an open submanifold of $\mathcal{N}$ with a piecewise $\mathcal{C}^1$-boundary $\dl \mathcal{U}$, then by the divergence theorem (see the appendix, Lemma \ref{divergencetheorem}) and the properties of $\mathbf{T}$, we have for $\eta$ a normal vector to $\dl \mathcal{U}$ and $\tau$ a transverse one such that $\eta^a \tau_a=1$:
\begin{eqnarray}
\int_{\dl \mathcal{U}} \mathbf{T}_{ab}X^b \eta^a i_{\tau} \d^4x &=& \int_{\mathcal{U}} \nabla^a\left(\mathbf{T}_{ab}X^b\right)  \d^4x \nonumber \\
&=&  \int_{\mathcal{U}} \left(X^b\nabla^a\mathbf{T}_{ab} + \mathbf{T}_{ab}\nabla^a X^b \right)  \d^4x \nonumber \\
&=&\hf\int_{\mathcal{U}} {}^{(X)}\pi_{ab} \mathbf{T}^{ab} \d^4x \; . \label{divergencetheoremSEtensor}
\end{eqnarray}

It is particularly interesting when $X$ is Killing and thus its deformation tensor vanishes. Motivated by this, we define the energy of a general $2$-form $F$ which the energy-momentum tensor depends on (aside from the metric), on an oriented smooth hypersurface $S$ to be:
\begin{equation}\label{enegrysurfacegeneralform}
E_X [F](S)=\int_S (X\hook \mathbf{T})^\sharp \hook \d^4 x \; ,
\end{equation}
or if we choose $\eta_S$ and $\tau_S$ to be respectively vector fields normal and  transverse to $S$, such that their scalar product is one\footnote{The existence of such vector fields follows from the fact that $S$ is a smooth orientable hypersurface of a smooth pseudo--Riemannian manifold.
}, it will be,
\begin{equation}\label{energyonasurface}
E_X [F](S)=\int_S \mathbf{T}_{ab}X^b \eta_S^a i_{\tau_S} \d^4x \; .
\end{equation}
Of course, it is understood that we are not integrating the $3$-form but its restriction on $S$, which is the pull back of the form by the inclusion map. If $S=\Sigma_t=\{t\}\times\R\times\mathcal{S}^2$, then its unit normal is $\hat{T}=f^{-\hf}\dl_t$, and taking the transverse vector to be $\hat{T}$ also, a simple calculation allows us to see that
\begin{eqnarray}
E_T[F](\Sigma_t)&=&\int_{\Sigma_t} \mathbf{T}_{ab}T^b \hat{T}^a i_{\hat{T}} \d^4x \nonumber \\
&=&\int_{\Sigma_t} \mathbf{T}_{00} f^{-\hf} f^{\hf}r^2 \d_{r_*} \d^2\omega \nonumber\\
&=&\frac{1}{4} \int_{\Sigma_t} |\Phi_1|^2 + \frac{2f}{r^2}|\Phi_0|^2 + |\Phi_{-1}|^2 \d r_* \d^2 \omega=E_T[F](t) \; .\label{energyofmaxwellont=cst}
\end{eqnarray}
This gives, 
by (\ref{divergencetheoremSEtensor}), that it is a conserved quantity as the vector field $T$ is Killing, i.e.
\begin{equation}\label{conservationofenergyMAXWELL}
E_T[F](t)=E_T[F](0) \; .
\end{equation}

Evidently not all solutions of Maxwell's equations decay in time, take for example the case where $\bm{\Phi}$ is a non zero constant vector, then it satisfies (\ref{NewmanPenrose}) and clearly does not decay as it does not change with time. Even solutions having finite energy, may not decay in time: Consider the constant vector $\bm{\Phi}=(\Phi_1=0~,~{\Phi}_0=C\neq0~,~{\Phi}_{-1}=0)$, it has finite energy, yet it does not decay. Since Maxwell's equations are linear, the last example shows that solutions  i.e. (even with finite energy) having charge do not decay, where by the charge or stationary part of a Maxwell field we mean the $l=0$ part of the spin--weighted spherical harmonic decomposition ( see $\Psi^0_{00}$ below). So, we need to exclude such solutions in order to have decay. 

%

In fact, and as known in the literature, it turns out that 

\begin{prop1}[Stationary Solutions]\label{stationarysolutionstheorem}
	The only admissible time--periodic solutions of Maxwell's equations with finite energy are exactly the pure charge solutions :
	\begin{equation}
	\mathbf{\Phi}=\left(
	\begin{array}{c}
	0 \\
	C \\
	0 \\
	\end{array}
	\right)                \qquad \textit{where C is a complex constant}.
	\end{equation}
\end{prop1}
\begin{proof}
	See \cite{mokdad_maxwell_2016,bachelot_gravitational_1991} for example.
\end{proof}

The solutions we will consider from now on are finite energy solutions with no stationary part, and by Proposition \ref{stationarysolutionstheorem}, these are the finite energy solutions in the orthogonal complement of the  $l=0$ subspace, that is solutions of the form:

\begin{eqnarray}
\Phi_{\pm 1}(t,r_*,\theta,\varphi)&=& \sum\limits_{l=1}^{+\infty}\sum\limits_{n=-l}^l\Psi_{\pm 1n}^l(t,r_*) W^l_{\pm 1n}(\theta,\varphi) \qquad\qquad \Psi_{\pm1n}^l  \in L^2(\R_{r_*})\; , \label{nonstationarysolutions1}\\
\Phi_0(t,r_*,\theta,\varphi)&=& \sum\limits_{l=1}^{+\infty}\sum\limits_{n=-l}^l\Psi_{0n}^l(t,r_*) W^l_{0n}(\theta,\varphi) \qquad\qquad \frac{\sqrt{f}}{r}\Psi_{0n}^l \in L^2(\R_{r_*})\; . \label{nonstationarysolutions}
\end{eqnarray}

where $$\{W^l_{mn}(\theta,\varphi) ;l,m,n \in \mathbb{Z}; l\geq0, -l\leq m,n\leq l \}$$ form an orthonormal basis of {spin-weighted spherical harmonics} of $L^2(\mathcal{S}^2)$.

In fact, if $F$ is a Maxwell field on $\mathcal{N}$ with spin components $\bm\Phi$ whose spin-weighted harmonic coefficients are $\bm{\Psi}^l_n$, then $F$ has a global potential if and only if the imaginary part of $\Psi^0_{00}$ vanish. Thus, as a consequence of the form (\ref{nonstationarysolutions1}) and (\ref{nonstationarysolutions}), the solutions we consider here have global potentials.

\section{Conformal Scattering}\label{sec:traceoper}

The first step in defining the scattering operator is to define the trace operators. 
We define the energy function spaces on the horizons and the initial Cauchy hypersurface, and we obtain an energy identity up to $i^+$. Then by the well--posedness of the Cauchy problem on the closure of $\mathcal{N}$, the trace operators are well defined.

\subsection{Function Space and Energy Identity}

Assume that $F$ is a smooth Maxwell field defined on $\bar{\mathcal{N}}$. The energy flux of the Maxwell field across an oriented hypersurface of $\bar{\mathcal{N}}$ is defined to be the quantity (\ref{energyonasurface}) with respect to the smooth vector field $T$ which is given by $\dl_t$ in the RNdS coordinate, and by $\dl_{u_\pm}$ on $\mathcal{M}^\pm$, and vanishes on the bifurcation spheres.

For any Cauchy hypersurface of constant $t$ the expression of the energy flux across it is given by (\ref{energyofmaxwellont=cst}). We therefore define the finite energy space $\mathcal{H}$ on $\Sigma:=\Sigma_0$ as the completion of the smooth compactly supported data consisting of triplets $(\Phi_{-1},\Phi_0,\Phi_1)\in\left(\mathcal{C}^\infty_0(\Sigma)\right)^3$. However, if we look at Maxwell's compacted equations and subtract (\ref{NewmanPenrose3}) from (\ref{NewmanPenrose2}), we see that we have a constraint equation on the spin--components that only involves spacial derivatives in directions tangent to $\Sigma$. It follows that in order for a triplet in $\mathcal{H}$ to be the initial data of a Maxwell field, i.e. the restriction of a Maxwell field on the Cauchy hypersurface, it must satisfy the constraint equation. Therefore, we need to restrict our Maxwell data to this constraint subspace of $\mathcal{H}$ which we will denote by $\mathcal{U}$. 

It is still possible to approximate data in $\mathcal{U}$ by smooth compactly supported data satisfying the constraint equation. In other words, $\left(\mathcal{C}^\infty_0(\Sigma)\right)^3\cap \mathcal{U}$ is dense in $\mathcal{U}$. One way to see this is by the fact that Maxwell equations can be reformulated in terms of a potential satisfying the Lorentz gauge condition as a hyperbolic system of four equations with four unknowns, without constraints. This means Maxwell data can always be approached by smooth compactly supported data. For details we refer to \cite{mokdad_maxwell_2016}. Since our spacetime is globally hyperbolic, then by Leray's theory for hyperbolic equations \cite{leray_hyperbolic_1955} we have:

\begin{prop1}\label{cauchyontheclosure}
	The Maxwell's Cauchy problem on $\bar{\mathcal{N}}$ is well-posed in $\mathcal{U}$, the constrained space of finite energy on $\Sigma_0$.
\end{prop1}

Across the horizons $\mathscr{H}^+_3$ and $\mathscr{H}^-_2$ the energy flux can be expressed using the outgoing tetrad defined above and the outgoing spin components in (\ref{spincomponentsofFinLhat}). Precisely, in the retarded coordinates $(u_-,r,\omega)$, $N=2\dl_{u_-}+f\dl_r$ is normal to these two horizons and is equal to $2T=2\dl_{u_-}$ on these null hypersurfaces. In addition, $\hat{L}=\dl_r$ is transverse to them and $g(\dl_{u_-},\dl_r)=1$. So, we take $\eta_{\mathscr{H}^+_3}=\hf N$ and $\tau_{\mathscr{H}^+_3}=\hat{L}$, and we have
\begin{equation*}
E_T[F]({\mathscr{H}^+_3})=\frac{1}{4}\int_{\mathscr{H}^+_3}\mathbf{T}_{ab}N^a N^b i_{\hat{L}} \d^4 x \; ,
\end{equation*}
which is,
\begin{equation*}
E_T[F]({\mathscr{H}^+_3})=-\frac{1}{4}\int_{\mathscr{H}^+_3}\ |\Phi_{-1}|^2 \d u_- \wedge \d^2 \omega \; ,
\end{equation*}
where we have chosen to orient ${\mathscr{H}^+_3}$ by $\dl_r$ so that $i_{\hat{L}} \d^4 x$ is a positively oriented volume form on it, and the above quantity is thus positive. In other words, $(\dl_{u_-},\dl_\theta,\dl_\varphi)$ is a negatively oriented frame on the horizon and so is the chart $(u_-,\omega)$, hence,
\begin{equation}\label{energyonH3+}
E_T[F]({\mathscr{H}^+_3})=\frac{1}{4}\int\limits_{\R_{u_-}\times\mathcal{S}^2}\ |\Phi_{-1}|^2 \d u_-  \d^2 \omega \; .
\end{equation}
The expression of $E_T[F]({\mathscr{H}^-_2})$ is exactly the same. As for the other two horizons ${\mathscr{H}^+_2}$ and ${\mathscr{H}^-_3}$ which are covered by the advanced coordinates $(u_+,r,\omega)$, we orient them by $\hat{N}=-\dl_r$ and use the incoming tetrad and the spin components analogous to (\ref{spincomponentsofFinLhat}), to have,
\begin{equation}\label{energyonH2+}
E_T[F]({\mathscr{H}^+_2})=\frac{1}{4}\int_{\mathscr{H}^+_2}\mathbf{T}_{ab}L^a L^b i_{\hat{N}} \d^4 x=\frac{1}{4}\int\limits_{\R_{u_+}\times\mathcal{S}^2}\ |\Phi_{1}|^2 \d u_+  \d^2 \omega \; ,
\end{equation}
and $E_T[F]({\mathscr{H}^-_3})$ has the same expression.

This gives us the definition of finite energy on the horizons $\mathscr{H}^{\pm}_i$. Compared to the expression of the energy flux (\ref{energyofmaxwellont=cst}) on a spacelike slice of constant $t$, we can almost see the conservation law up to the horizons: If we take ``limits'' as $t$ goes to $\pm\infty$, the surface $\Sigma_t$ approaches $\mathscr{H}^\pm_2 \cup \mathscr{H}^\pm_3$ respectively, and since $f=0$ on the horizons, we formally have 
$$\lim_{t \rightarrow \pm \infty} E_T[F](\Sigma_t) = E_T[F]({\mathscr{H}^\pm_2})+E_T[F]({\mathscr{H}^\pm_3}),$$
but because of the energy conservation in (\ref{conservationofenergyMAXWELL}), one expects the following conservation law 
\begin{equation}\label{conservationlawuptoi+}
E_T[F](\Sigma_0) = E_T[F]({\mathscr{H}^\pm_2})+E_T[F]({\mathscr{H}^\pm_3}).
\end{equation}

Thus, we define the energy spaces on the horizons $\mathscr{H}^\pm_i$ to be the completions of $\mathcal{C}^{\infty}_0(\mathscr{H}^\pm_i)$ with respect to the the norms
\begin{equation}\label{normonONEhorizon}
\Vert \phi \Vert_{\mathscr{H}^\pm_2}^2 =\pm \frac{1}{2}\int_{\mathscr{H}^{\pm}_2}\ |\phi|^2 \d u_\pm \wedge \d^2 \omega \; , \quad ; \quad \Vert \phi \Vert_{\mathscr{H}^\pm_3}^2 =\mp \frac{1}{2}\int_{\mathscr{H}^{\pm}_3}\ |\phi|^2 \d u_\mp \wedge \d^2 \omega \; .
\end{equation}
On the future and past {total horizons}\index{Total horizon} $\mathscr{H}^\pm :=\mathscr{H}^\pm_2 \cup \mathscr{H}^\pm_3$\index{$\mathscr{H}^\pm$}, we define the energy space $\mathcal{H}^\pm$ to be the completions of $ \mathcal{C}^{\infty}_0(\mathscr{H}^\pm_2)\times \mathcal{C}^{\infty}_0(\mathscr{H}^\pm_3)$ with respect to the addition norm 
\begin{equation}\label{normonTWOhorizons}
\Vert (\phi_\pm,\phi_\mp) \Vert_{\mathcal{H}^\pm}^2=\frac{1}{2}\Vert \phi_\pm \Vert^2_{\mathscr{H}^\pm_2}+\frac{1}{2}\Vert\phi_\mp \Vert_{\mathscr{H}^\pm_3}^2.
\end{equation}
\begin{figure}
	\centering
	\includegraphics[scale=1]{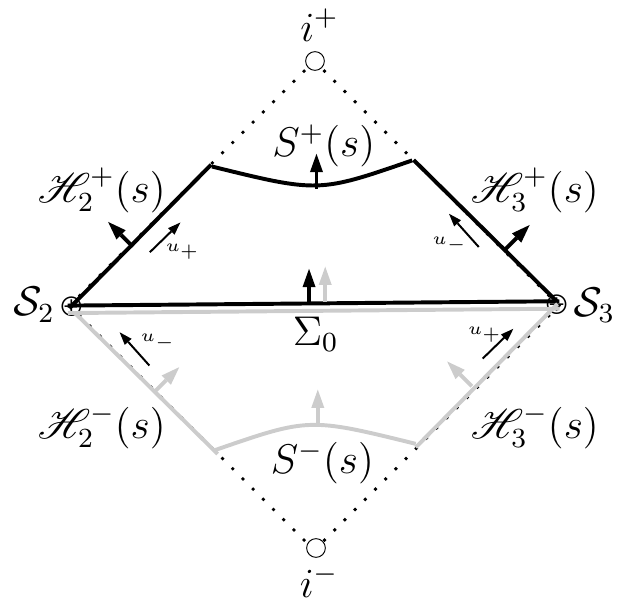}
	\caption{\textit{The hypersurfaces $S^\pm (s)$ forming two closed hypersurfaces (black and gray). The thick arrows indicate the orientation of the surface, while the thin arrows indicate the direction of increasing coordinate $u_\pm$ from $-\infty$ to $+\infty$.}}
	\label{fig:energyconservSurfaces}
\end{figure}
We now use the decay results obtained in \cite{mokdad_decay_2017} (Theorem 19) to show that these norms are conserved for smooth compactly supported data. Consider the hypersurfaces 
\begin{equation}\label{hyperparabolas}
S^\pm (s)=\{(t,r_* ,\omega)\in \R \times \R \times \mathcal{S}^2 \; ; ~t=\pm \sqrt{1+r_*^2} + s ~;~ \pm s \ge 0\}.
\end{equation}
$S^+(s)$ actually intersects $\mathscr{H}^+$ in two spheres, one in each of the horizons $\mathscr{H}^+_2$ and $\mathscr{H}^+_3$ , namely at $$\{ s \}_{u_+} \times \{ r_2 \}_r \times \mathcal{S}^2~~ \mathrm{and}~~ \{ s\}_{u_-}\times\{r_3\}_r \times \mathcal{S}^2$$ respectively (figure \ref{fig:energyconservSurfaces}). Therefore if we set 
\begin{eqnarray*}
	\mathscr{H}^+_2(s)&=&\ ]-\infty , s[_{ u_+}\times\{r_{2}\}_r \times \mathcal{S}^2\; ,\\ 
	\mathscr{H}^+_3(s)&=&]-\infty, s[_{ u_-}\times\{r_{3}\}_r \times \mathcal{S}^2\; ,
\end{eqnarray*}
then these hypersurfaces along with $\Sigma_0$ and $S^+(s)$, in addition to the bifurcation spheres $\mathcal{S}_2$ and $\mathcal{S}_3$,  form a closed hypersurface in $\bar{\mathcal{N}}$. The same goes for $S^-(s)$, as shown in figure \ref{fig:energyconservSurfaces}. Thus, if $F$ is a smooth solution of Maxwell's equations which is compactly supported for each $t$, then since $T$ is Killing on $\bar{\mathcal{N}}$, we have by (\ref{divergencetheoremSEtensor}),
\begin{equation*}
E_T[F](\Sigma_0) = E_T[F]({\mathscr{H}^+_2(s)})+E_T[F]({\mathscr{H}^+_3}(s))+E_T[F](S^+(s)).
\end{equation*}
$E_T[F]({\mathscr{H}^+_2(s)})$ and $E_T[F]({\mathscr{H}^+_3}(s))$ are two positive increasing functions of $s$, and from the positiveness of $E_T[F](S^+(s))$, their sum is bounded from above by $E_T[F](\Sigma_0)$. Thus they have limits when $s$ tends to $+\infty$, and these limits are $E_T[F]({\mathscr{H}^+_2})$ and $E_T[F]({\mathscr{H}^+_3})$. Thanks to uniform decay proved in \cite{mokdad_decay_2017}, 
$$\lim_{s\rightarrow + \infty} E_T[F](S^+(s))=0 ,$$
and the conservation law (\ref{conservationlawuptoi+}) is proved. The same holds true with past horizons and $S^-(s)$.

\subsection{Trace Operators}

\begin{figure}
	\centering
	\includegraphics[scale=1]{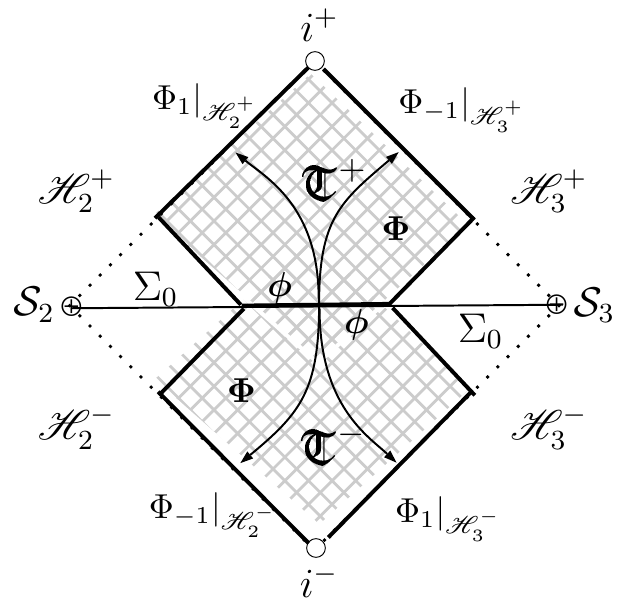}
	\caption{\textit{The Trace operators defined for smooth compactly supported Maxwell Cauchy data.}}
	\label{fig:Traceoperators}
\end{figure}
We now define the trace operators. Since we only showed that the energy is conserved for smooth fields with compact supports for each $t$, we first define the future and past trace operators by
\[\textgoth{T}^\pm : \left(\mathcal{C}^\infty_0(\Sigma)\right)^3 \cap \mathcal{U}\longrightarrow \mathcal{H}^\pm ,\]
and as follows:
Let $F_0$ be the 2-form on $\Sigma_0$ whose spin components in the stationary tetrad $\{L,N,M,\bar{M}\}$ are the initial Cauchy data $\bm{\phi}=(\phi_1,\phi_0,\phi_{-1}) \in \left(\mathcal{C}^\infty_0(\Sigma)\right)^3 \cap \mathcal{U}$, and let $\tilde{F}$ be the unique solution to the Cauchy problem on $\bar{\mathcal{N}}$ with $\tilde{F}|_{\Sigma_0}=F_0$, and whose spin components in the tetrad $\{L,N,M,\bar{M}\}$ are $\bm{\Phi}=(\Phi_1,\Phi_0,\Phi_{-1})$, then
$$\textgoth{T}^\pm (\bm{\phi})=(\Phi_{\pm1}|_{\mathscr{H}^\pm_2},\Phi_{\mp1}|_{\mathscr{H}^\pm_3}),$$
as figure \ref{fig:Traceoperators} illustrates, and by (\ref{conservationlawuptoi+})
\begin{equation}\label{traceconservenorm}
\Vert\bm{\phi} \Vert_{\mathcal{H}}=\Vert \textgoth{T}^\pm(\bm{\phi})\Vert_{\mathcal{H}^\pm}.
\end{equation}

By the density of $\left(\mathcal{C}^\infty_0(\Sigma)\right)^3\cap \mathcal{U}$ in $\mathcal{U}$, $\textgoth{T}^{\pm}$ extend to bounded operators on $\mathcal{U}$: $$\textgoth{T}^\pm :  \mathcal{U}\longrightarrow \mathcal{H}^\pm ,$$ 
with closed range, and still satisfy (\ref{traceconservenorm}).


The main result of this section is that the trace operators defined above are invertible and hence isometries, allowing us to introduce the scattering operator. Since an isometry is a surjective norm preserving linear map between Hilbert spaces, all that is left is to show that the trace operators are surjective.

More precisely, let $(\phi_\pm , \phi_\mp)\in \mathcal{H}^\pm$, we wish to show that there exist some Cauchy data $\bm{\phi}\in \mathcal{U}$ such that $\textgoth{T}^\pm(\bm{\phi})=(\phi_\pm , \phi_\mp)$, and since the trace operators are injective, the Cauchy data $\bm{\phi}$ is unique if it exists. By the well-posedness of the Cauchy problem, this will mean that there is a unique finite energy solution $\bm{\Phi}$ such that $\bm{\Phi}|_{\Sigma_0}=\bm{\phi}$, and by the definition of $\textgoth{T}^\pm$ we have  
$$\textgoth{T}^\pm (\bm{\Phi}|_{\Sigma_0})=(\Phi_{\pm1}|_{\mathscr{H}^\pm_2},\Phi_{\mp1}|_{\mathscr{H}^\pm_3}),$$
and hence  
\begin{equation}\label{Goursatinitialcondition}
(\Phi_{\pm1}|_{\mathscr{H}^\pm_2},\Phi_{\mp1}|_{\mathscr{H}^\pm_3})=(\phi_\pm , \phi_\mp).
\end{equation}
Therefore, what we want to do is to solve the characteristic Cauchy problem, also known as the Goursat problem, on the total horizons $\mathscr{H}^\pm$. We do so by showing that the ranges of the trace operators contain dense subsets of the Hilbert spaces $\mathcal{H}^{\pm}$, and since norm preserving linear maps take complete normed spaces to complete ones, this means that the ranges are equal to $\mathcal{H}^{\pm}$. Thus, by density, it is enough to consider {Goursat data}\index{Goursat data(Maxwell)} in $\mathcal{C}^\infty_0\left(\mathscr{H}^\pm_2\right)\times \mathcal{C}^\infty_0\left(\mathscr{H}^\pm_3\right)$. As the future and the past cases are analogous, we only work out the case of the future trace operator. To further simplify the problem, we take advantage of the linearity of Maxwell's equations and assume that the non-trivial part of the initial (Goursat) data is only on one horizon, i.e. we treat smooth compactly supported data of the form, say, $(0,\phi_-)\in \mathcal{H}^+$ with $\phi_-\in\mathcal{C}^\infty_0\left(\mathscr{H}^+_3\right) $, which represent the trace of an outgoing Maxwell solution. The case of $(\phi_+,0)\in \mathcal{H}^+$ is completely analogous.

\subsection{Goursat Problem and the Scattering Operator}\label{goursatproblemsection}

To solve the Goursat problem we use the results of L. H\"{o}rmander \cite{hormander_remark_1990} by first converting the initial-value problem from Maxwell's equations to wave equations, then following J.-P. Nicolas \cite{nicolas_conformal_2016} in his approach of putting the problem in a framework for which H\"{o}rmander's results apply. The idea is then to reinterpret the solution of the wave equations obtained, using the results of \cite{hormander_remark_1990}, as a Maxwell field. 

As we shall restrict our attention to the future cosmological horizon, let us consider the outgoing tetrad and the corresponding spin components of the Maxwell field in details. If $F$ is a Maxwell field, then  its spin components 
satisfy coupled wave equations. In particular, this is true for the outgoing components in $\bm{\hat{\Phi}}=(\hat{\Phi}_1,\Phi_0,\Phi_{-1})$. 

\begin{lem1}\label{waveequationoutgoingLEMMA}
	Let $\bm{\hat{\Phi}}=(\hat{\Phi}_1,\Phi_0,\Phi_{-1})$ be the outgoing spin components of a smooth Maxwell field defined on $\mathcal{M}^-_F$, then $\bm{\hat{\Phi}}$ satisfies the wave equation 
	\begin{equation}\label{waveequationoutgoingtetrad}
	\hat{W}\bm{\hat{\Phi}}=\left(\begin{array}{ccc}
	\hat{W}_{11} & - \hat{V}'M & 0 \\
	0 &\hat{W}_{00}& 0 \\
	0 & - V'\bar{M}& \hat{W}_{0-1}
	\end{array}\right)\left(\begin{array}{c}
	\hat{\Phi}_1\\
	\Phi_0\\
	\Phi_{-1}
	\end{array}\right)=0,
	\end{equation}
	where differentiation with respect to $r$ is indicated by a prime, $N_1=N-f' \;$, and the diagonal entries are\footnote{\index{$\hat{W}_{ij}$}The indices of $\hat{W}_{ij}$ indicate their expressions: $\hat{W}_{ij}=\hat{L}\;I(N)-\hat{V}\;J(M,M_1)$ with
		\begin{align*}
		i=\begin{cases}
		0 & \mathrm{if}~ I(N)=N\; ;\\
		1 & \mathrm{if}~ I(N)=N_1\; ,
		\end{cases} 
		&&
		j=\begin{cases}
		1& \mathrm{if}~ J(M,M_1)=M\bar{M}_1 \; ;\\
		0 & \mathrm{if}~ J(M,M_1)=M_1\bar{M}=\bar{M}_1 M\; ;\\
		-1 & \mathrm{if}~ J(M,M_1)=\bar{M}M_1 \; .
		\end{cases} 
		\end{align*}}
	\begin{align}
	\hat{W}_{11}:=\hat{L}N_1 - \hat{V}M\bar{M}_1\; , && \hat{W}_{00}:=\hat{L}N- \hat{V}M_1\bar{M}\; , && \hat{W}_{0-1}:=\hat{L}N - \hat{V}\bar{M}M_1 \;.
	\end{align}
\end{lem1}
\begin{proof}
	We denote the left hand side of Maxwell's equations (\ref{CEsoutgoing1})-(\ref{CEsoutgoing4}) as:
	\begin{eqnarray}
	N_1\hat{\Phi}_1 - \hat{V} M\Phi_0&=:&E_1  \; ;  \label{E1}\\
	\hat{L}\Phi_0 - \bar{M}_1\hat{\Phi}_1&=:&E_2 \; ; \label{E2}\\
	N\Phi_0  + M_1\Phi_{-1}&=:&E_3\; ;  \label{E3}\\
	\hat{L}\Phi_{-1}  +\hat{V}\bar{M}\Phi_0&=:&E_4\; . \label{E4}
	\end{eqnarray}
	
	As $[N,\hat{L}]=f'\hat{L}$, i.e. $N_1\hat{L}=\hat{L}N$, and $M_1\bar{M}=\bar{M}_1 M$, we have:
	\begin{eqnarray}
	\hat{L}E_1 +\hat{V}M E_2&=&\hat{W}_{11}\hat{\Phi}_1  - \hat{V}'M \Phi_0 \; ; \label{W11} \\
	N_1E_2 + \bar{M}_1 E_1&=&\hat{W}_{00} \Phi_0 \; ;\label{W00E12}\\
	\hat{L}E_3-M_1 E_4&=&\hat{W}_{00} \Phi_0  \; ; \label{W00E34}\\
	N_1E_4-\hat{V}\bar{M}E_3 &=&\hat{W}_{0-1}\Phi_{-1} -V'\bar{M}\Phi_0\; .\label{W0-1}
	\end{eqnarray}
	Finally, to see that $\hat{W}$ is indeed a modified d'Alembertian we just note that 
	$$\hat{L}N-\hat{V}M_1\bar{M}=\square + r\hat{V}(f\hat{L}-N),$$
	where 
	\begin{equation}\label{waveequation}
	\square =\square_g =\nabla^\alpha \nabla_\alpha =g^{\mathbf{ab}}(\dl_\mathbf{a}\dl_\mathbf{b}
	- \Gamma^\mathbf{c}_\mathbf{ab}\dl_\mathbf{c}),
	\end{equation}
	is the d'Alembertian of the geometric wave eqaution.
\end{proof}

We now look at Maxwell's equations on $\mathscr{H}^+_3$, in particular the first and the third, $E_1=0$ and $E_2=0$. Since $N$ is tangent to the horizon, equations (\ref{CEsoutgoing1}) and (\ref{CEsoutgoing3}) are tangent to it, i.e. contain only tangential derivatives:
\begin{eqnarray}
{N_1}|_{\mathscr{H}^+_3} {\hat{\Phi}_1}|_{\mathscr{H}^+_3} -\hat{V}(r_3)M {\Phi_0}|_{\mathscr{H}^+_3}&=&0  \; , \label{constraintsonhorizon1}\\
N|_{\mathscr{H}^+_3} \Phi_0 |_{\mathscr{H}^+_3}  + M_1\Phi_{-1}|_{\mathscr{H}^+_3}&=&0 \; . \label{constraintsonhorizon2}
\end{eqnarray} 
These are the constraints on the horizon. Thus, they must be satisfied by the restriction of the field's spin components. It follows that if the Goursat data $\phi_-\in\mathcal{C}^\infty_0\left(\mathscr{H}^+_3\right)$ is to be viewed as part of a Maxwell field, namely $\Phi_{-1}|_{\mathscr{H}^+_3}$, then the other two components of the field are determined uniquely on the horizon by $\phi_-$ through the above constraints and the requirement that they vanish in a neighbourhood of $i^+$. This is because (\ref{constraintsonhorizon1}) and (\ref{constraintsonhorizon2}) force them to vanish identically from $i^+$ to the support of the Goursat data $\phi_-$. We choose them to be zero near $i^+$ since as we shall presently see, this allows us to apply Hörmander's result. Therefore, for $\phi_-\in\mathcal{C}^\infty_0\left(\mathscr{H}^+_3\right)$ we define $\phi_0,\hat{\phi}_+\in \mathcal{C}^\infty\left(\mathscr{H}^+_3\right)$ consecutively by the constraints initial-value problems in $\mathscr{H}^+_3$:
\begin{equation}\label{spincomponentsonH3+}
(C_1)\begin{cases}
2\dl_{u_-} \phi_0= M_1 \phi_- \\
\phi_0|_{\mathcal{S}_p}=0
\end{cases}
\qquad ; \qquad\qquad
(C_2)\begin{cases}
(2\dl_{u_-}-f'(r_3)) \hat{\phi}_+= \hat{V}(r_3) M \phi_0 \\
\hat{\phi}_+|_{\mathcal{S}_p}=0
\end{cases}
\end{equation}
where $\mathcal{S}_p$ is any sphere of $\mathscr{H}^+_3$ in the future of the support of $\phi_-$. The supports of $\phi_0$ and $\hat{\phi}_+$ may touch the bifurcation sphere $\mathcal{S}_3$, but this is no problem since $\mathcal{S}_3$ is a finite smooth sphere in $\bar{\mathcal{N}}$ where the Cauchy hypersurface meets the future cosmological horizon, and no real scattering happens at $\mathcal{S}_3$. We refer to the triplet 
\begin{equation*}\label{Goursatdataforwave}
\bm{\hat{\phi}}=(\hat{\phi}_+,\phi_0,\phi_-)
\end{equation*}
also as the Goursat data, since it will be the {Goursat data}\index{Goursat data (Wave)} for the wave equations (\ref{waveequationoutgoingtetrad}).

In \cite{hormander_remark_1990}, the author consider Lorentzian manifolds with a time function whose level hypersurfaces are compact and spacelike. The work is actually done for product manifolds of the form $\R\times X$ where $X$ is smooth compact manifold without boundary on which a time dependent Riemannian metric is defined, and the Laplace-Beltrami operator is defined with respect to a fixed Riemannian density. The paper studies the well-posedness of the Cauchy problem set on weakly spacelike hypersurfaces that are the graphs of Lipschitz functions over $X$, for wave equations of the form:
\begin{equation}\label{hormanderwaveequation}
\square u+ Q u = h \; ,
\end{equation}
where $\square$ is a the modified d'Alembertian while $Q$ is a first order operator of essentially bounded measurable coefficients, and $h$ a source. In fact, H\"{o}rmander's results are valid for globally hyperbolic and spatially compact spacetimes, since the product structure can be recovered by global hyperbolicity, while any non-degenerate change in the metric or the volume density, entails in the d'Alembertian a change that can be absorbed into the first order operator $Q$.  

In what comes next, we need the well-posedness of the Goursat problem on the future total horizon for different wave equations (see Lemma \ref{waveequationoutgoingLEMMA}) that are of the form (\ref{hormanderwaveequation}). And although our spacetime is not spatially compact (without boundary), as long as the Goursat data is smooth and supported away form $i^+$, then its compact support in $\mathscr{H}^+_3 \cup \mathcal{S}_3$ enables us to transform the problem into a framework suitable for H\"{o}rmander's results. Following the work of J.-P. Nicolas \cite{nicolas_conformal_2016}, this is done through the following construction.

\begin{figure}
	\centering
	\includegraphics[width=\textwidth]{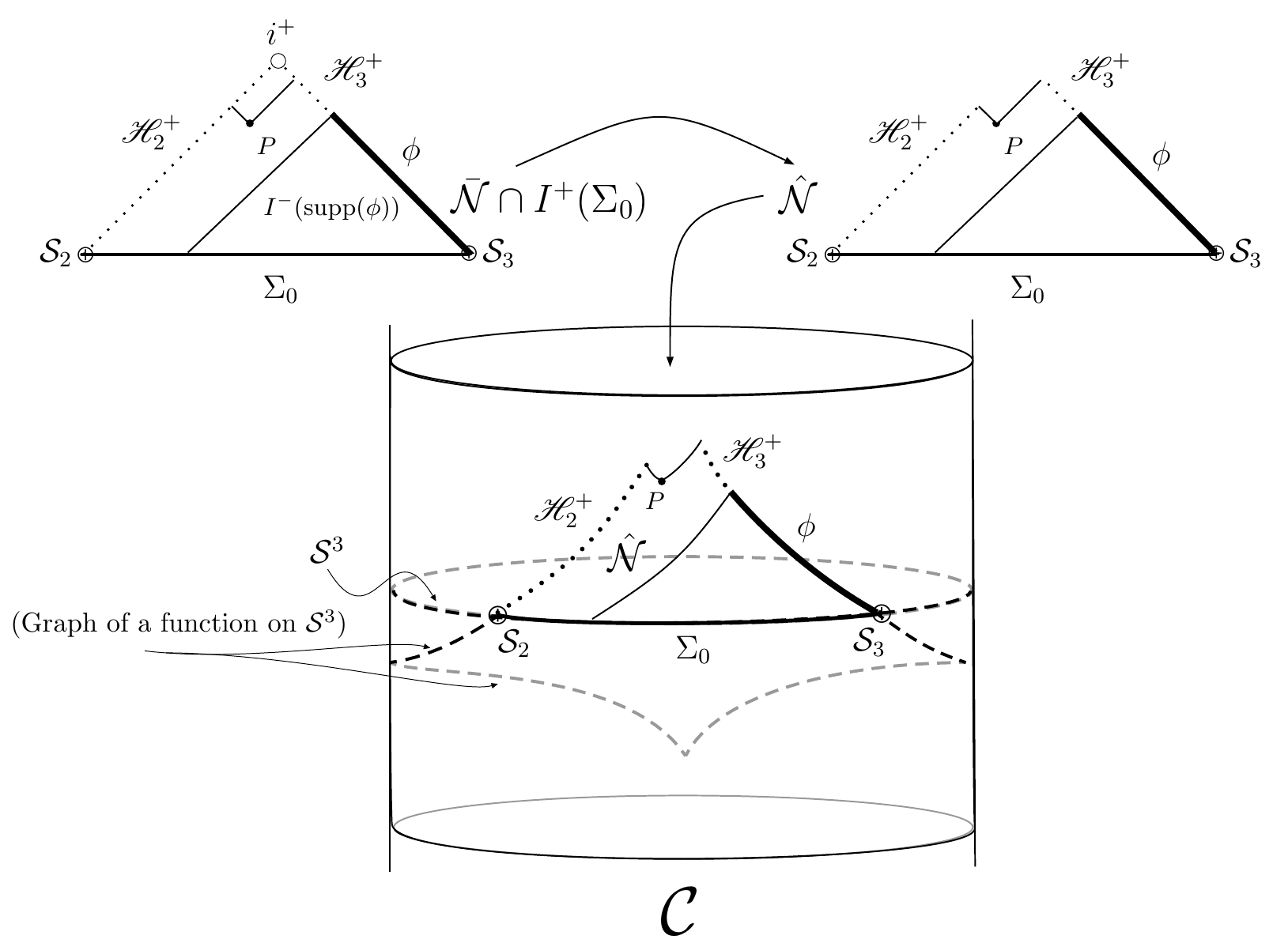}
	\caption{\textit{The construction done to understand the Goursat problem for data $\phi$ supported away from $i^+$ in a framework suited to H\"{o}rmander's result.}}
	\label{fig:goursatextend}
\end{figure}

We pick any point $P$ whose future does not intersect the support of the Goursat data on the horizon, that is, a point in the future of the past of the data. We then remove the future of this point, and set $$\hat{\mathcal{N}}=\left(\bar{\mathcal{N}} \setminus I^+(p)\right)\cap I^+(\Sigma_0),$$ where $I^+(\Sigma_0)$ is the future of the Cauchy hypersurface $\Sigma_0$ in $\bar{\mathcal{N}}$. We now extend $\hat{\mathcal{N}}$ as a globally hyperbolic cylindrical spacetime $(\mathcal{C}=\R \times \mathcal{S}^3,\tilde{g})$. We extend $\Sigma_0$ as $\mathcal{S}^3$ and the remaining part of the future total horizon as the graph of a Lipschitz function over $\mathcal{S}^3$, and the data by zero on the rest of the extended hypersurface. Then \cite{hormander_remark_1990} guarantees the existence of a unique smooth solution on $\mathcal{C}$ to the wave equation we consider. We take the restriction of the solution to $\hat{\mathcal{N}}$. Finite propagation speed then ensures that the solution is zero in the future of the past of the Goursat data (figure \ref{fig:goursatextend}). 

Moreover, despite the fact that (\ref{waveequationoutgoingtetrad}) is a coupled system of three scalar wave equations, the coupling happens only on lower order terms, meaning that $\square$ is in the diagonal only. Thus, the work in \cite{hormander_remark_1990}, where a single scalar wave equation (not a system) with scalar source is treated, can be applied to our case, when put in the above framework, with only a slight modification\footnote{The general operator $Q$ of first and lower order terms in the equation considered in \cite{hormander_remark_1990} is controlled by a priori estimates giving exponential bounds in \cite{hormander_remark_1990}. If the lower order term is a matrix instead of a simple scalar potential, it can be controlled in the same manner, and the proof goes through unchanged.}. However in truth, the results in \cite{hormander_remark_1990} can be applied to (\ref{waveequationoutgoingtetrad}) directly and without any modification at all since (\ref{waveequationoutgoingtetrad}) can be considered as three separate single scalar wave equations, two of which have a source, and one is source-free. This is because the middle component, $\Phi_0$, satisfies the decoupled source-free wave equation  $\hat{W}_{00}\Phi_0=0$, and the coupling is only between the middle component and each of the other components separately. Hence the terms depending on $\Phi_0$ in the other two equations can simply be viewed as source terms after solving $\hat{W}_{00}\Phi_0=0$.

\begin{thm1}[Goursat Problem]\label{GourastproblemTHEOREM}
	For $\phi_-\in\mathcal{C}^\infty_0\left(\mathscr{H}^+_3\right) $ there is a unique smooth, finite energy, Maxwell field $F$ defined on $\bar{\mathcal{N}}$, with ${\bm{{\Phi}}}=({\Phi}_1,\Phi_0,\Phi_{-1})$ its spin components in the stationary tetrad, such that 
	$$(\Phi_{1}|_{\mathscr{H}^+_2},\Phi_{-1}|_{\mathscr{H}^+_3})=(0 , \phi_-).$$
\end{thm1}
\begin{proof}
	Finite energy is immediate from the law of conservation of energy (\ref{conservationlawuptoi+}), and thus uniqueness follows directly from the injectivity of the future trace operator and the well-posedness of the Cauchy problem on $\bar{\mathcal{N}}$.
	
	Let  $\phi_0$ and $\hat{\phi}_+$ be given by $\phi_-$ and (\ref{spincomponentsonH3+}), so that 
	\begin{align*}
	N_1 |_{\mathscr{H}^+_3} {\hat{\phi}_+} -\hat{V}(r_3)M {\phi_0}&=0  \; , \tag{a} \label{constrainta}\\
	N|_{\mathscr{H}^+_3} \phi_0   + M_1\phi_{-}&=0 \; , \tag{b} \label{constraintb}
	\end{align*}
	and set $\bm{\hat{\phi}}=(\hat{\phi}_+,\phi_0,\phi_-)$. We now extend  $\bm{\hat{\phi}}$ by zero to $\mathscr{H}^+_2$. The reason we do so, is because H\"{o}rmander's results apply to Goursat data defined on a generalized Cauchy hypersurface\footnote{Weakly spacelike hypersurface such that every inextendible timelike curve intersect it only once.}, so we consider our data to be defined on the future total horizon $\mathscr{H}^+$. By \cite{hormander_remark_1990} there is a unique smooth solution $\bm{\hat{\Phi}}=(\hat{\Phi}_1,\Phi_0,\Phi_{-1})$ to the Goursat problem
	\begin{equation}\label{Goursatproblemofwaveequation}
	\begin{cases}
	\hat{W}\bm{\hat{\Phi}}=0 \\
	\bm{\hat{\Phi}}|_{\mathscr{H}^+}=\bm{\hat{\phi}}
	\end{cases}
	\end{equation}
	\begin{figure}
		\centering
		\includegraphics[scale=1]{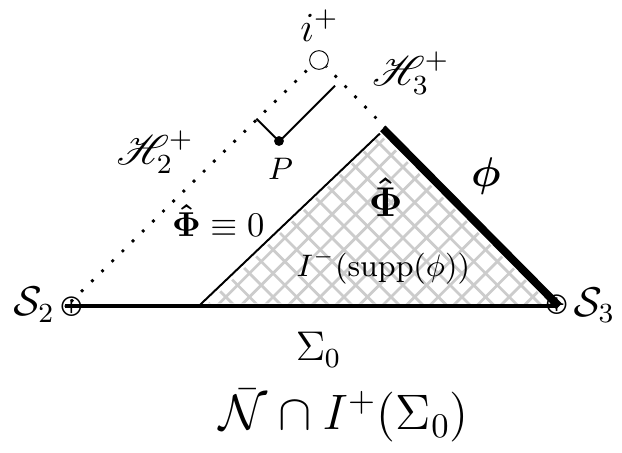}
		\caption{\textit{The solution of the wave equations and its support.}}
		\label{fig:solutionsupport}
	\end{figure}
	defined on $\bar{\mathcal{N}}\cap\left(\Sigma_0\cup I^+(\Sigma_0)\right)$. A nd by finite propagation speed and local uniqueness, $\bm{\Phi}$ is zero on $I^+(I^-(\mathrm{supp}\bm{\phi}))$, of course except for the part of the 
	horizon where the support lies (figure \ref{fig:solutionsupport}).
	
	We now reinterpret $\bm{\hat{\Phi}}$ as the spin components\footnote{Although the outgoing tetrad is singular on $\mathscr{H}^+_2$, $\bm{\hat{\Phi}}$ vanishes on a neighbourhood of $\mathscr{H}^+_2$, so we can take our generalized Cauchy hypersurface to be $\mathscr{H}^+_3\cup S'$, where $S'$ is a null hypersurface in $I^+(I^-(\mathrm{supp}\bm{\phi}))$  as in figure \ref{fig:cauchyE4}.} of a solution to the Goursat problem on $\mathscr{H}^+$ for Maxwell's equations with data $(0,\phi_-)$. 
	Let $\hat{W}\bm{\hat{\Phi}}=(\hat{\Omega}_1,\Omega_0 ,\Omega_{-1})$, then using (\ref{W11})-(\ref{W0-1}) only, we have
	\begin{align}
	N_1 \hat{\Omega}_1 - \hat{V}M\Omega_0&=\hat{W}_{01}E_1+f\hat{V}'ME_2 \; ;\label{Omega1}\\
	\hat{L}{\Omega}_0 - \hat{V}\bar{M}_1\hat{\Omega}_1&=\hat{W}_{10}E_2 \; ; \label{Omega2}\\
	N_1{\Omega}_0 +{M}_1{\Omega}_{-1}&=\hat{W}_{00}E_3 \; ;\label{Omega3}\\
	\hat{L}{\Omega}_{-1} + \hat{V}\bar{M}\Omega_0&=\hat{W}_{1-1}E_4-\hat{V}'ME_3 \; . \label{Omega4}
	\end{align}
	where 
	\begin{align}
	\hat{W}_{01}&=\hat{L}N - \hat{V}M\bar{M}_1 \; ;\\
	\hat{W}_{10}&=\hat{L}N_1 - \hat{V}M_1\bar{M} \; ;\\
	\hat{W}_{1-1}&=\hat{L}N_1 - \hat{V}\bar{M}_1{M} \; .
	\end{align}
	Since (\ref{Goursatproblemofwaveequation}) holds, then on the one hand, we see that the $E_i$'s are solutions of coupled wave equations, and on the other hand, the constraints (\ref{constrainta}) and (\ref{constraintb}) implies that $E_1|_{\mathscr{H}^+}=0$ and $E_3|_{\mathscr{H}^+}=0$. 
	It follows that $E_3$ is a solution of the Goursat problem 
	\begin{equation*}
	\begin{cases}
	\hat{W}_{00}E_3=0 \\
	E_3|_{\mathscr{H}^+}=0
	\end{cases}
	\end{equation*}
	and hence $E_3=0$. This has an immediate effect on $E_4$ by (\ref{W0-1}), i.e. $N_1 E_4=0$, and in particular, we now have $N_1|_{\mathscr{H}^+} E_4|_{\mathscr{H}^+}=0$.
	But since $\bm{\hat{\Phi}}$ is zero in a neighbourhood of $i^+$ which intersects the horizon, all the derivatives of its components vanish as well, among which are $\hat{L}\Phi_{-1}$ and $\bar{M}\Phi_0$. (\ref{E4}) then means that $E_4|_{\mathscr{H}^+}=0$, and therefore $E_4$ solves the Goursat problem
	\begin{equation*}
	\begin{cases}
	\hat{W}_{1-1}E_4=0 \\
	E_4|_{\mathscr{H}^+}=0
	\end{cases}
	\end{equation*}
	and so $E_4=0$. 
	\begin{figure}
		\centering
		\includegraphics[scale=1]{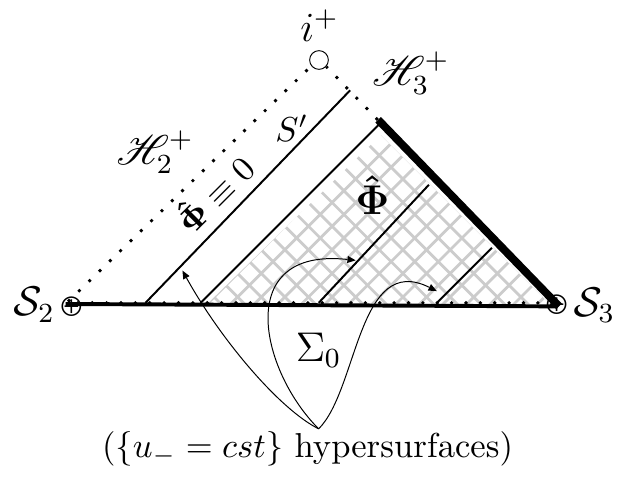}
		\caption{\textit{The foliation by the hypersurfaces $\{u_-=cst\}$ for the equation $N_1 E_4=0$.}}
		\label{fig:cauchyE4}
	\end{figure}
	We could have alternatively used equation $N_1 E_4=0$ directly to show that $E_4=0$: Because of the form of supp$(\bm{\hat{\Phi}})$, $\hat{L}\Phi_{-1}$ and $\bar{M}\Phi_0$ are zero on a hypersurface $S'$ of constant $u_-$ lying in the future of $I^-(\mathrm{supp}\bm{\phi})$, and so $E_4|_{S'}=0$ (figure \ref{fig:cauchyE4}). Now for the simple transport equation $N_1 E_4=0$, the initial-value problem 
	\begin{equation*}
	\begin{cases}
	N_1 E_4=0 \\
	E_4|_{S'}=0
	\end{cases}
	\end{equation*}
	is well-posed and has a unique solution, thus, $E_4=0$.
	
	Because only $E_1$ is tangential to the horizon while $E_2$ is the one satisfying a source-free wave equation among the two, we need to use both at the same time. The fact that $E_1|_{{\mathscr{H}^+}}=0$ implies that $N_1|_{\mathscr{H}^+} E_2|_{\mathscr{H}^+}=0$ by (\ref{W00E12}) and the fact that $N$ is tangent to the horizon. Now by the above argument of zero derivatives near $i^+$, $E_2|_{\mathscr{H}^+}$ itself is zero on some sphere at the horizon, say $\mathcal{S}_p$. Therefore $E_2 |_{\mathscr{H}^+}$ in turn solves 
	\begin{equation*}
	\begin{cases}
	N_1|_{\mathscr{H}^+} E_2|_{\mathscr{H}^+}=0 \\
	(E_2|_{\mathscr{H}^+})|_{\mathcal{S}_p}=0
	\end{cases}
	\end{equation*}
	which is a well-posed initial-value problem on the 2-\emph{surface} $\mathcal{S}_p$ \emph{in} the horizon. Thus, $E_2|_{\mathscr{H}^+}=0$, and so,
	\begin{equation*}
	\begin{cases}
	\hat{W}_{10}E_2=0 \\
	E_2|_{\mathscr{H}^+}=0
	\end{cases}
	\end{equation*}
	i.e. $E_2=0$. For $E_1$, we now have two options, both follow from what we have so far. Either we consider $E_1$ as the solution of the Goursat problem
	\begin{equation*}
	\begin{cases}
	\hat{W}_{01}E_1=0 \\
	E_1|_{\mathscr{H}^+}=0
	\end{cases}
	\end{equation*}
	\begin{figure}
		\centering
		\includegraphics[scale=1]{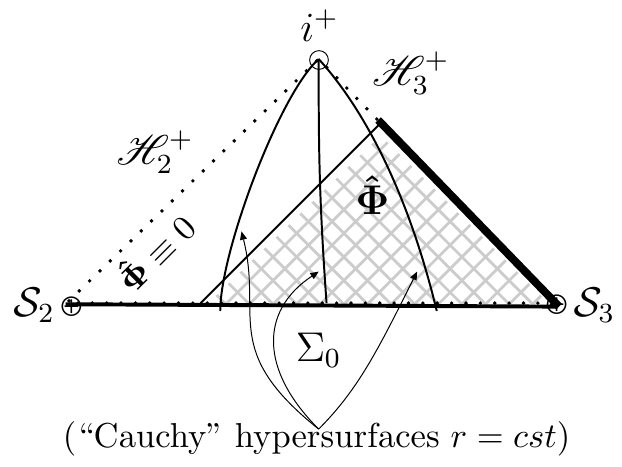}
		\caption{\textit{The foliation by the hypersurfaces $\{r=cst\}$ for the equation $\hat{L} E_1=0$.}}
		\label{fig:cauchyE1}
	\end{figure}
	where the initial condition is given by the constraint (\ref{constrainta}), or, we use (\ref{W11}) as a simple initial-value problem 
	\begin{equation*}
	\begin{cases}
	\hat{L}E_1=0 \\
	E_1|_{\mathscr{H}^+_3}=0
	\end{cases}
	\end{equation*}
	and since ${\mathscr{H}^+_3}$ is a hypersurface of constant $r$, the problem is well-posed (figure \ref{fig:cauchyE1}). Both methods entails that $E_1=0$.
	
	Therefore, $\bm{\hat{\Phi}}$ are the outgoing components of a Maxwell field $F$. The well-posedness of the Cauchy problem on $\Sigma_0$ for Maxwell's equations ensures the global definition of $F$ on $\bar{\mathcal{N}}$ as a smooth solution. The only thing left to prove is that $F$ has zero trace on the future outer horizon, which, since $F$ is smooth up to the horizons, follows from the relation $f\hat{\Phi}_1=\Phi_1$ and the fact that $\hat{\Phi}_1$ vanishes on a neighbourhood of the future outer horizon $\mathscr{H}^+_2$.
\end{proof}

\subsubsection*{Scattering Operator}

Theorem \ref{GourastproblemTHEOREM} shows that the trace operators $\textgoth{T}^\pm$ have inverses and are in effect isometries from $\mathcal{U}$ to $\mathcal{H}^\pm$.

The scattering operator is the map $\textgoth{S}:\mathcal{H}^-\longrightarrow\mathcal{H}^+$ defined as follows: 
\begin{equation*}\label{TheHolyGrail}
\textgoth{S}=\textgoth{T}^+\circ(\textgoth{T}^-)^{-1} 
\end{equation*}

\begin{appendices}

\section{Divergence Theorem}

One important tool that we shall use is the divergence theorem. We present a version of this theorem which we think is better suited for Lorentzian geometry than the usual one used for on Riemannian geometry. 
Let $\mathcal{U}$ be an oriented smooth $n$-manifold. Fix $\omega$ a positively oriented volume form , i.e. determining the orientation on $\mathcal{U}$, and let $X$ be a smooth vector field on it. 
The divergence of $X$ is defined to be the function $divX$ such that,
\begin{equation}\label{divergenceinternsicexpression}
\lie_X\omega=(divX)\omega \; .
\end{equation}
If the orientation on $\mathcal{U}$ is given by a pseudo-Riemannian metric $g$, then we can choose $\omega=\d V_g$, and the above definition of $divX$ coincides with the more familiar one, which is locally defined as:
\begin{equation}\label{divergencelocalexp1}
\frac{1}{\sqrt{|g|}}\dl_i \left(\sqrt{|g|}X^i\right)\; ,
\end{equation}
where $|g|$ is the absolute value of the determinant of the metric $g$.

\begin{lem1}[Divergence Theorem]\label{divergencetheorem}
	Let $\mathcal{U}$ be an oriented smooth n-manifold with boundary (possibly empty), with $\omega$ a positively oriented volume form , i.e. determining the orientation on $\mathcal{U}$, and the boundary $\dl \mathcal{U}$ is outward oriented (Stokes' orientation), and let $X$ be a smooth vector field on $\mathcal{U}$. If $\mathcal{U}$ is compact or $X$ is compactly supported then,
	
	\begin{equation}\label{divergencetheoremgeneralformula}
	\int_{\dl \mathcal{U}} i_X \omega = \int_{\mathcal{U}}  \lie_X \omega= \int_{\mathcal{U}}  divX \omega \; ,
	\end{equation}
	
	Moreover, if the orientation on $\mathcal{U}$ is given by a pseudo-Riemannian metric $g$, i.e. $\omega=\d V_g$, then (\ref{divergencetheoremgeneralformula}) can be reformulated as:
	
	\begin{equation}\label{divergencetheoremnormal-transverseformula}
	\int_{\dl \mathcal{U}} N (X)  i_L \d V_g = \int_{\mathcal{U}}   \lie_X \d V_g = \int_{\mathcal{U}}   divX \d V_g\; ,
	\end{equation}
 	where $N$ is a conormal field to $\dl \mathcal{U}$, i.e. $N^\sharp$ is a normal vector field, and $L$ is a vector field transverse (nowhere tangent) to $\dl \mathcal{U}$, such that $N(L)=1$.
\end{lem1}

If the normal vector field can be normalized (which is always the case if the metric is Riemannian and is  true in the Lorentzian case only if the hypersurface is timelike), one can then choose the transverse vector to be the normal itself and thus recovering the well known form of this theorem:
\begin{equation}\label{divergencetheoremwellknownformula}
\int_{\dl \mathcal{U}} N(X) \d V_{\tilde{g}} = \int_{\mathcal{U}} div X \d V_g \; , \hspace{1cm} \textrm{or,}  \hspace{1cm} \int_{\dl \mathcal{U}} N_a X^a \d V_{\tilde{g}} = \int_{\mathcal{U}} \nabla_a X^a \d V_g
\end{equation}
$\tilde{g}$ being the induced metric on $\dl \mathcal{U}$, and $\d V_{\tilde{g}}=i_{N^\sharp}\d V_g$.

Killing vector fields have the nice property of vanishing divergence. A vector field $X$ is said to be Killing if the metric is conserved along the flow of $X$, i.e. $\lie_X g=0$. 
Thus, $\lie_X g_{ab}=\nabla_a X_b + \nabla_b X_a= 2\nabla_{(a} X_{b)}$, and for Killing fields,
\begin{equation}\label{killingequation}
\nabla_a X_b - \nabla_b X_a=0 \; ,
\end{equation}
consequently,
\begin{equation}\label{killingvanishdivergence}
0=g_{ab}\left(\nabla_a X_b - \nabla_b X_a\right)=2\nabla_a X^a \; ,
\end{equation}
hence $divX=0$. Equation (\ref{killingequation}) is called the Killing equation, and the $(0,2)$-tensor involved is sometimes called the deformation tensor or Killing tensor, denoted
\begin{equation}\label{deformationtensor}
{}^X \pi_{ab}= 2\nabla_{(a} X_{b)} \; .
\end{equation}

%
%

\end{appendices}

\section*{Acknowledgement}
The results of this paper, the mentioned decay results, and the study of the RNDS spacetimes \cite{mokdad_decay_2017,mokdad_reissner-nordstrom-sitter_2017}, were obtained during my PhD thesis \cite{mokdad_maxwell_2016}. I would like to thank my thesis advisor Pr. Jean-Philippe Nicolas for his indispensable guidance during the thesis.

\printbibliography[heading=bibintoc]

\end{document}